\documentclass[5p,times]{elsarticle}

\usepackage{ecrc}

\usepackage{amsmath}
\usepackage{graphicx}
\usepackage{units}
\usepackage{color}
\usepackage{lipsum}
\usepackage{multirow}
\usepackage{amssymb}
\usepackage[figuresright]{rotating}
\usepackage{picins}
\usepackage{hyperref}

\usepackage[pscoord]{eso-pic}

\newcommand{\placetextbox}[3]{
  \setbox0=\hbox{#3}
  \AddToShipoutPictureFG*{
    \put(\LenToUnit{#1\paperwidth},\LenToUnit{#2\paperheight}){\vtop{{\null}\makebox[0pt][c]{#3}}}%
  }%
}%

\DeclareMathOperator*{\E}{\large \textrm{E}}

\newcommand{\xv}{\mbox{\boldmath$x$}}
\newcommand{\zv}{\mbox{\boldmath$z$}}

\newcommand{\yv}{\mbox{\boldmath$y$}}

\newcommand{\etav}{\mbox{\boldmath$\eta$}}
\newcommand{\xiv}{\mbox{\boldmath$\xi$}}

\newcommand{\fv}{\mbox{\boldmath$f$}}
\newcommand{\gv}{\mbox{\boldmath$g$}}
\newcommand{\nuv}{\mbox{\boldmath$\nu$}}
\newcommand{\Iv}{\mbox{\boldmath$I$}}
\newcommand{\hv}{\mbox{\boldmath$h$}}

\newcommand{\wv}{\mbox{\boldmath$w$}}
\newcommand{\varrhov}{\mbox{\boldmath$\varrho$}}

\volume{00}
\firstpage{1}
\journalname{Ad Hoc Networks}
\runauth{M. Mongelli, S. Scanzio}
\jid{procs}
\jnltitlelogo{Ad Hoc Networks}
\CopyrightLine{2016}{Published by Elsevier Ltd.}

\begin{document}

\placetextbox{0.5}{1}{This is the author's version of an article that has been published in this journal.}
\placetextbox{0.5}{0.985}{Changes were made to this version by the publisher prior to publication.}
\placetextbox{0.5}{0.97}{The final version of record is available at \href{https://doi.org/10.1016/j.adhoc.2016.06.002}{https://doi.org/10.1016/j.adhoc.2016.06.002}}%

\begin{frontmatter}

\dochead{}

\title{A neural approach to synchronization in wireless networks with heterogeneous sources of noise}

\author{Maurizio Mongelli\corref{cor1}\fnref{label3}}
\ead{maurizio.mongelli@ieiit.cnr.it}
\author{Stefano Scanzio\fnref{label4}}
\cortext[cor1]{Corresponding author.}
\ead{stefano.scanzio@ieiit.cnr.it}

\fntext[label3]{Area della Ricerca di Genova, via De Marini 6, Genova, Italy.}
\fntext[label4]{Corso Duca degli Abruzzi 24, 10129 Torino, Italy.}

\address{Institute of Electronics, Computer and Telecommunication
Engineering, National Research Council of Italy}

\begin{abstract}
The paper addresses state estimation for clock synchronization in the presence of factors affecting the quality of synchronization. Examples are temperature variations and delay asymmetry. These working conditions make synchronization a challenging problem in many wireless environments, such as Wireless Sensor Networks or WiFi. Dynamic state estimation is investigated as it is essential to overcome non-stationary noises. The two-way timing message exchange synchronization protocol has been taken as a reference. No a-priori assumptions are made on the stochastic environments and no temperature measurement is executed. The algorithms are unequivocally specified offline, without the need of tuning some parameters in dependence of the working conditions. The presented approach reveals to be robust to a large set of temperature variations, different delay distributions and levels of asymmetry in the transmission path.
\end{abstract}

\begin{keyword}
Clock synchronization protocols \sep Dynamic state estimation \sep Two-way timing message exchange \sep WSN \sep WiFi
\end{keyword}

\end{frontmatter}

\section{Introduction} \label{sec:intro}
Clock Synchronization Protocols (CSPs) have a fundamental role in many technological contexts in which a common time reference is required \cite{2013-IEM-CSP1}.
For example, synchronization is used in Wireless Sensor Networks (WSNs) \cite{AdHoc-intro-sync-WSN}, localization \cite{AdHoc2, AdHoc-intro-sync-localization}, home automation \cite{intro-sync-AVB}, industrial networks \cite{TII-sync-industrial}, traffic scheduling \cite{AdHoc-intro-traffic-scheduling, Ber15}, and in a number of other contexts in which actuation and/or sensing must be synchronous.
The quantities measuring the asynchronism between the clocks of two nodes in a network are: the \emph{offset}, i.e., the difference between the two clocks and the \emph{skew}, i.e., the normalized difference between the Crystal Oscillator (XO) oscillation frequency and its nominal frequency. The variable component of the skew is the \emph{drift}. Their precise estimation defines the target of the CSP and they are jointly optimized \cite{AdHoc3}. They typically represent the state of the synchronization problem, when it is formulated under dynamic state equations.

The estimation process may be severely compromised by a number of factors. The most important are: the random delays affecting the communication path between nodes, including software or hardware delays inside them, the precision of nodes in timestamping events, and changes in the environment conditions.
Estimation of offset and skew may be driven by signal processing techniques, which assume a time-fixed state (see, e.g., \cite{magazine} for WSNs) or by dynamic state tracking, e.g., through Kalman filtering (as an example, see \cite{ToIM} for IEEE 1588 protocol). Addressing time-varying conditions means to follow instantaneous fluctuations due to non-stationary noises, such as temperature variations of XOs \cite{2004-TUTORIAL-VIG, temperature, temperature3}.

\subsection{Background and objectives} \label{sec:backgroundandobjectives}
In the present paper, we study how to compensate with a single technique all the possible factors affecting the synchronization quality.
The idea to analyze and compensate a number of causes together is not new.
Algorithms derived from machine learning (e.g., neural networks, support vector machines,...) are typically exploited to model complex processes, in the case a theoretical model is not known or it cannot be parameterized in practice because too many measurements of the real system are needed for a satisfactory characterization. The latter is the case of synchronization protocols. For example, the oscillation frequency of an XO is influenced by several environmental factor: the temperature, the supply voltage, vibrations, age, etc. All these factors, well documented in the scientific literature \cite{2004-TUTORIAL-VIG}, have not the same influence on the behavior of different types of XOs, and even the same type of XOs differently reacts to environmental conditions, depending on its manufacturing process. As a consequence, to compensate all these factors, each XO must be experimentally characterized with respect to the physical phenomena that can modify its behavior. Such a kind of analysis can only be performed during the manufacturing process of the component, because XOs are usually soldered on the motherboard. All these physical quantities must then be sensed at runtime for their relevant compensation. This last step is not easy and often it is not feasible because, for example, XOs do not usually include temperature sensors. In this case, the estimation must be derived by using sensors in the proximity of the XO. On the other hand, XOs that automatically compensate some external factors exist, but they are hardly integrated in commercial devices since they cost too much. The same difficulties in finding a correct model apply also to other quantities such as the timestamps precision and accuracy, and asymmetric delays. They depend on the hardware, but in the case of software timestamps also on the interference caused by other processes executed in the operating system of the node. For these reasons, we decide to focus on an algorithm that compensates all these aspects together. Some results concentrate the attention on temperature, because it is the most affecting environmental factor.

The algorithm we pursue should be capable to work with minimal online adjustment of the parameters, thus avoiding the need of reconfigurations following the actual behavior of the noises.

Synchronization is formulated as a dynamic state tracking problem beyond regular LQG hypoteses\footnote{Linear dynamics of the system, quadratic cost function and Gaussian noises.} because temperature measurement noise may not always be Gaussian in practical systems \cite{temperature, temperature3, nostroToIM}. The inherent optimal estimation filter may be hardly derived in closed form. This approach typically has consequences in terms of numerical analysis with complex operations (see, e.g., the Particle Filtering in \cite{magazine}), which are not easily applicable in devices in which computational power or energy are scarce resources. Since the investigated suboptimal filter is based upon neural approximation, the approach may lead to a heavy computational effort in the offline phase (during which the training of the neural network is provided), but synchronization corrections are provided online almost instantly.
Delay assymetry is also addressed jointly with temperature variations. This avoids configuring countermeasures to assymetry that are separate from the rest of the synchronization scheme.

\subsection{Contribution} \label{sec:contribution}
The method firstly outlined in \cite{nostroToIM} for \textit{receiver-receiver} CSPs \cite{TII-sync-industrial}, is now applied in the sender-receiver context, more used in practice, and under realistic conditions of WSN and WiFi networks, including delay asymmetry. Despite the considered CSP drives delay compensation, we show how no knowledge of delay is necessary for the used estimation techniques. An enhancement of the method is proposed to cope with exponentially distributed delays, a condition not often detectable in practice, but analyzed in some scientific works \cite{ICCSvedesi}. The method provides good generalization capabilities to different delays distributions (i.e., Gaussian and exponential delays). The multi-hop context is also addressed to limit computational cost and simplify the applicability of the method.

\subsection{Organization of the paper}
The paper is organized as follows. The next section deals with the analysis of the state of the art and highlights the position of the present paper. Section \ref{sec:problem} addresses the mathematical formulation of the estimation problem.
The subsequent sections enter in the details of the estimation techniques proposed, including computational and implementation aspects. Section \ref{sec:experimentssetting} defines the setting of the experiments and Section \ref{sec:performance} discusses the results. Conclusions and future work are finally outlined at the end of the paper.

\section{Related literature} \label{sec:literature}

\subsection{State estimation}
Dynamic state estimation for synchronization is an open issue for environments with non-Gaussian and non-stationary noises \cite{temperature, temperature3}. An example for WSN has been reported in \cite{magazine}, by introducing Particle Filtering (PF). \cite{magazine} shows how addressing time-varying conditions may considerably improve the synchronization gain over signal processing techniques. PF is able to adapt to Gamma distributed delays better than signal processing, which works well under Gaussian or exponential delays. PF belongs to the optimal Bayesian framework for dynamic state estimation. This is exactly the research line we want to pursue here, without incurring in the computational burden involved by PF.

As far as signalling processing techniques are concerned, our approach has been compared with \cite{Chaudhari}, which is a reference target in this field (see, e.g., \cite{ICCSvedesi}), since it presents a computationally light approach, which is also robust to the underlying network delay density function and asymmetry. More refined techniques are available as well, for example, in the presence of exponentially distributed delays \cite{ICCSvedesi}.

\subsection{Parameters setting}
Online adaptation may be critical if the statistical parameters of the noises cannot be known in advance. More specifically, the covariance matrix of the noises is typically used as a parameter of the mentioned algorithms (Kalman \cite{ToIM}, signal processing as in \cite{ICCSvedesi} and Particle Filtering (PF) in \cite{magazine}). How parameters setting may be a critical task in Kalman is evidenced by \cite{EKalman}, in which practical guidelines are provided. This critical aspect has been also registered by \cite{temperature3}, in which the parameters of the estimation algorithm are tuned online and by \cite{temperature}, in which the parameters of the temperature-skew mapping are supposed to be known in advance. Synchronization solutions with self-learning capabilities may be hardly found in the literature. \cite{AdHoc1} has recently investigated how to adapt the time window of linear regression. The approach has been tested in stationary Gaussian conditions.

\subsection{Temperature noise}
Recent works address synchronization in WSNs by overcoming the temperature noise. In \cite{temperature}, the thermal drift is removed in advance, by exploiting the relationship between XO frequency and the temperature. A multi-model Kalman filter is studied in \cite{temperature3} to obtain the model likelihood for the skew, based on the measured temperature. The main advantage of the two approaches relies on the possibility to reduce the sending rate of synchronization messages, by keeping unchanged the synchronization quality since the temperature is locally compensated. An ARMAX model is studied in \cite{ToUltrasonic} to compensate temperature and aging effects. An upper bound of the error is derived in closed-form under Gaussian assumptions. The mentioned works rely on a mapping table from temperatures to clock skews \cite{Xu16}. \cite{Xu16} models the correlation between clock skews and temperature variations through the least squares method, thus achieving more flexibility, still relying on temperature measurements. The approach presented here does not exploit any measurement of the temperature. \cite{temperature2} deals with high latency networks by introducing a new message exchange in two steps: in the first one the delay is estimated and, in the second one, Kalman is applied. The refined procedure reveals to be robust to noise, including temperature changes.

\subsection{Asymmetric Delays}
Despite \cite{Chaudhari, ICCSvedesi} do not address delay asymmetry explicitly, they reveal to be robust to several working conditions, including asymmetry. More recent works address the mitigation of delay asymmetry \cite{AsyLv, Lee1, Lee2, Asy1, Asy2} explicitly. Timestamping corrections are provided to compensate the synchronization error induced by asymmetry. \cite{AsyLv} requires additional messages in the protocol. \cite{Lee1, Lee2} exploit different kinds of link speed measurements to infer the level of asymmetry. As evidenced in \cite{Asy1}, those measurements may be not always sufficient if the internal delays of the device have a predominant role. Proper statistical information is derived from additional link/internal device delays \cite{Asy1}. In the very rare case of intermediate devices without the compensation of the packet residence time inside the device, traffic queues can be measured \cite{Asy2}. The inherent corrections in \cite{Asy1} may require an accurate setup of the devices. \cite{ISPCS14} and \cite{Techno1} apply the Boot-strap method under the assumption of Gamma and exponential distributed bias in asymmetry, respectively. A similar approach is applied to Pareto distributed delays in \cite{Techno2}. The robustness of the methods are accurately analyzed with respect to parameters of the probability distributions. An important advantage of \cite{ISPCS14} consists of the simple calculations executed to derive the bias estimation. In \cite{Asy2}, the corrections may be sensitive to parameters changes (e.g., size of the observation window) and an accurate analysis is needed for them. Here, the correction is derived without any additional measurements or knowledge of the device and it is applied jointly to the rest of the compensation steps.

Although in practice the communication channel is sufficiently symmetric for the majority of the applications to not affect too much the synchronization quality, this consideration does not hold for in-node latencies, i.e., the delay inside the nodes between the sending/reception of a packet and the acquisition of the relevant timestamp exploited in clock correction. This problem also applies to such nodes that acquire the timestamp in hardware, and its effect on synchronization quality is clearly amplified if the network contains heterogeneous nodes. From the viewpoint of the synchronization protocols, in-node or communication channel asymmetries are indistinguishable and they have exactly the same consequence on the achievable synchronization quality. This evidence will be analyzed in detail in subsection \ref{sub:innode_delays}, which is based on data derived from scientific literature and acquired from real devices.

\subsection{Position of the paper}

\begin{table}[t]
\caption{Topics of research in synchronization and state of the art.}
\label{tab:tablecontribution}
\begin{center}
\begin{tabular}{l|c|c|c|c}

              & temp. & asym. & unknown & real    \\
              &       &       & noises  & applic. \\ \hline \hline
\cite{AdHoc3}      &          &           &           & $\bullet$    \\
\hline
\cite{ToIM}        &          &           &            & $\circ$     \\
\hline
\cite{temperature} & $\circ$  &           &            & $\bullet$   \\
\hline
\cite{temperature3}& $\circ$  &           &            &             \\
\hline
\cite{Xu16}        & $\circ$  &           & $\bullet$  & $\circ$     \\
\hline
\cite{Chaudhari}   &          & $\circ$   & $\bullet$  & $\bullet$   \\
\hline
\cite{AdHoc1}      &          &           & $\bullet$  &             \\
\hline
\cite{nostroToIM}  & $\bullet$&           & $\bullet$  & $\circ$     \\
\hline
\cite{ICCSvedesi}  &          & $\circ$   &            & $\bullet$   \\
\hline
\cite{EKalman}     &          &           & $\circ$    & $\circ$     \\
\hline
\cite{temperature2}& $\bullet$&           &            & $\bullet$   \\
\hline
[20, 23--29]       &          & $\bullet$ & $\circ$    & $\circ$     \\
\hline
Present paper      & $\bullet$& $\bullet$ & $\bullet$ & $\circ$      \\
\hline
\end{tabular}
\end{center}
\end{table}

Table \ref{tab:tablecontribution} summarizes the discussion presented in this section and highlights the contribution of the present paper. A $\bullet$ mark is assigned if the paper exactly addresses the topic of interest. A $\circ$ mark is assigned if the topic is partially addressed; for example, a $\circ$ mark is assigned to the mechanisms based on temperature measurements. As summarized by the table, the aims of the present paper are partially matched by the current literature.
The table also includes another important topic: the applicability of the algorithm in a real context. This pertains computational cost and ease of implementation and it is archived by algorithms requiring simple mathematical operations (such as summations, multiplications); a topical example is the one of \cite{Chaudhari} or the application of linear regression  \cite{AdHoc1}. The papers highlighted with a $\bullet$ mark on applicability hardly match the other requirements. The present approach may require a computational expensive training phase. We consider such a training phase the necessary step to achieve a good compromise between performance and applicability when adaptation to unknown noises, temperature and asymmetry compensations are required. Elaborating a countermeasure to those factors on the basis of samples of the system in a single algorithm is the topical issue addressed in this work. For this reason the paper is presented in the table (last row) with all $\bullet$ marks, except for the $\circ$ mark on applicability in virtue of the computational complexity of training. Similar considerations may hold for the use of the least square method of \cite{Xu16} or for the multi-model Kalman filter in \cite{temperature3} and PF in \cite{magazine}. Another point of strength of the present work is the performance metric used (the $99.9$ percentile of the synchronization error), which is even more stringent than the $90$\% and $92$\% confidence intervals of average absolute error of \cite{Zom14} and \cite{Xu16}, respectively.

\section{Problem formulation} \label{sec:problem}

\subsection{The two-way timing message exchange}
\begin{figure}[t]
\hskip-0.5in
\centering
\includegraphics[width=3in]{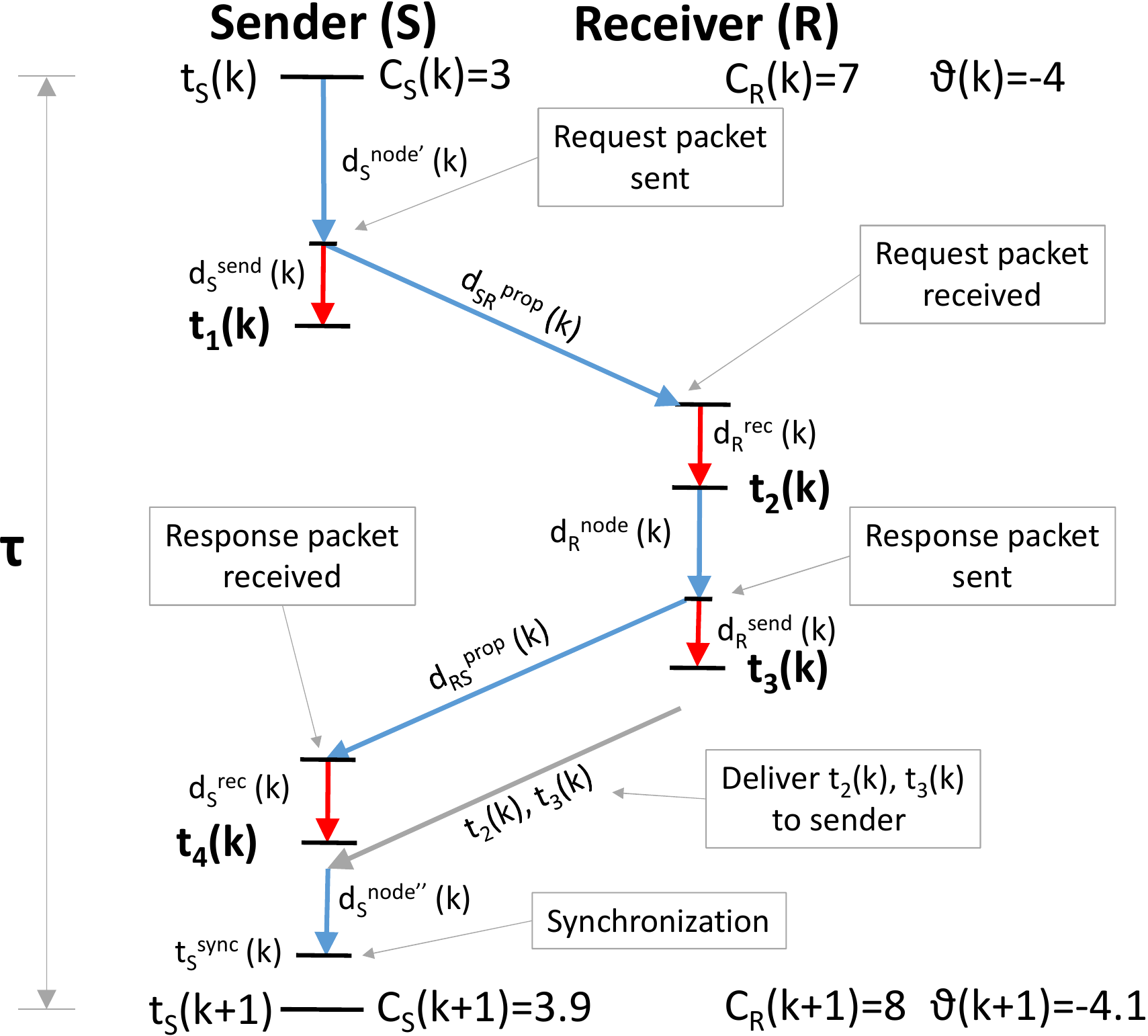}
\caption{Timestamps, messages exchange, and involved delays.} \label{fig:fig_delays-crop}
\end{figure}

We are mostly considering WSN and WiFi networks by focusing on pairwise synchronization (synchronization between a pair of neighboring nodes) rather than network-wide synchronization (hierarchical pairwise synchronization) \cite{magazine}; the network-wide model is a generalization of the pairwise model as outlined in \cite{ICCSvedesi}. We take the \textit{two-way timing message exchange} mechanism as a reference. This basic synchronization scheme is typical of many \textit{sender-receiver} CSPs, such as, e.g., the Timing-sync Protocol for Sensor Networks (TPSN) \cite{TPSN} and the \textit{timing measurement} mechanism defined in the recent IEEE 802.11-2012 specification of WiFi \cite{80211-2012}. We consider two nodes, called \emph{sender} and \emph{receiver}, which periodically take and exchange timestamps of their internal clocks. The sender is the one starting the exchange that consists of $3$ packets. On sending and reception of the first $2$ packets, $4$ timestamps are acquired, denoted by $t_1, t_2, t_3, t_4$ (see Fig.~\ref{fig:fig_delays-crop}, in which the timestamps are reported in \textbf{bold}). $t_1$ is the sending time of a synchronization packet from the sender to the receiver (under the notion of time of the sender). $t_2$ is the time of the receiving of the packet at the receiver and $t_3$ is the time of the sending of the response synchronization packet from the receiver to the sender. Both $t_2$ and $t_3$ are defined under the notion of time of the receiver. The last packet (from the receiver to the sender) includes the values of $t_2$ and $t_3$. Finally, $t_4$ is the time of the receiving of the first response packet (under the notion of time of the sender). The involved delays evidenced by Fig.~\ref{fig:fig_delays-crop} will be detailed later. At the end of the exchange, a measure of the synchronization state (offset, skew, together with the delay) is obtained. In turn, the state estimation is updated from the collected measurements at the end of each exchange; typically a set of $K$ message exchanges is exploited for state estimation \cite{Chaudhari, ICCSvedesi}. The sender makes use of the state estimation to synchronize its clock to that of the receiver. A software layer, named for the first time \textit{virtual clock} in \cite{2000-SRDS-virtualclock}, converts the sender time into the receiver time. A virtual clock is essential for devices that cannot adjust the clock register at runtime. Synchronization is performed at each discrete time instant $k, k+1, ...$; let $\tau$ be the size of those discrete time steps. We assume each message exchange starts and ends in $[k, k+1], \forall k$.

We now enter in the details of the state and measurement models.

\subsection{State equations}\label{sub:state_equations}
The clock registers of the sender and the receiver (denoted by $C_S(k)$ and $C_R(k)$, respectively and reported in Fig.~\ref{fig:fig_delays-crop}) differ of the offset quantity $\theta(k)$.
\begin{eqnarray}
C_S(k) = C_R(k)+\theta(k) \label{eq:CSCM}
\end{eqnarray}
A typical clock model can be represented by the following equations in the discrete domain:
\begin{eqnarray}
\theta(k) & = & \theta(k-1) + \gamma(k-1) \cdot \tau + \omega_{\theta}(k-1) \label{eq:state1}\\
\gamma(k) & = & \gamma(k-1) + \omega_{\gamma}(k-1) \label{eq:state2}\\
d(k) & = & d(k-1) \label{eq:state3}
\end{eqnarray}
where $\gamma(k)$ represents the skew and $d(k)$ the delay component, which is assumed stationary over time, but whose measurements are affected by noise as detailed later in subsection \ref{sec:measurementnoise}. The stationarity assumption is motivated in the same section as well.

\subsection{State noise}\label{sub:state_noise}
The $\omega_{\theta}$ and $\omega_{\gamma}$ quantities represent the noises affecting $\theta$ and $\gamma$, respectively, whose distributions are usually modeled as Gaussian type; the corresponding standard deviations are denoted by $\sigma_{\omega_{\theta}}$ and $\sigma_{\omega_{\gamma}}$, respectively.

Here and in \cite{nostroToIM} we include in $\omega_{\gamma}$ an additional component,
\begin{eqnarray}
\omega_{\gamma^{T}}(t,\cdot) & = &\omega_{\gamma^{T}}(t,tc,p,T^{High}_E,T^{Low}_E) \label{eq:noise-temp}
\end{eqnarray}
to mimic the temperature effect of a XO periodically moved every $p$ [s] between two environments with temperatures $T^{High}_E$ [${^{\circ}C}$] and $T^{Low}_E$ [${^{\circ}C}$]; $tc$ is the XO \textit{thermal time constant}, which represents the thermal inertia of the XO and of its case.

The \textit{Newton's law of cooling}
\begin{eqnarray}
T = T_E+(T_{XO}-T_E) \cdot e^{-\frac{1}{tc}\Delta t} \label{eq:newton}
\end{eqnarray}
models the evolution over time of the temperature of an object with initial temperature $T_{XO}$ which is placed in an environment characterized by a new temperature $T_E$. The periodic movement between the two environments with different temperatures has been performed in simulation every $p$, and in such instants $T_{XO}$ is set equal to the current temperature $T$ of the XO and $T_E$ is set equal to $T^{High}_E$ or $T^{Low}_E$.

The XO is modeled as an AT-cut quartz, the more common in real devices and the one that suffers more the temperature effects. Given the temperature $T$, the frequency variation from the nominal one can be properly approximated as
\begin{eqnarray}
\frac{\Delta f}{f} = a\cdot(T-T_0) + b\cdot(T-T_0)^2 + c\cdot(T-T_0)^3 \label{eq:tempFreqChar}
\end{eqnarray}
where $T_0$ is the reference temperature of the XO and $a$, $b$, $c$ are three constants modeling the XO. This paper makes use of parameters directly derived from a real AT-cut quartz \cite{1962-IRE-XO}: $T_0=\unit[25]{^{\circ}C}$, $a=0.0$, $b=0.4 \cdot 10^{-9}$ and $c=109.5 \cdot 10^{-12}$.

The quantity $\omega_{\gamma^{T}}(t,\cdot)$, which corresponds to the frequency variation $\frac{\Delta f}{f}$ at a given time $t$, can be easily derived by substituting (\ref{eq:newton}) in (\ref{eq:tempFreqChar}).

We noticed that the temperature variations have a predominant effect on the synchronization error rather than other variations of the parameters in (\ref{eq:tempFreqChar}) ($a$, $b$, $c$, $T_0$). For this reason, the performance evaluation concentrates the attention on large temperature variations under a realistic XO model \cite{1962-IRE-XO}, with fixed parameters.

The inherent probability distribution of $\omega_{\gamma^{T}}(t,\cdot)$ is not Gaussian; it is actually a multi-modal distribution, with significant asymmetry among the peaks.
\subsection{Measurement equations} \label{sec:measurementequations}
Let $t_1(k), t_2(k), t_3(k), t_4(k)$ the timestamps in $[k-1, k]$. A sample of the delay at time $k$, $\check d(k)$, is derived as follows:
\begin{eqnarray}
\check d(k) = \frac{ \Big(1-\gamma(k)\Big)\cdot\Big(t_4(k)-t_1(k)\Big)-\Big(t_3(k)-t_2(k)\Big) }{2} \label{eq:measure1}
\end{eqnarray}
The equation, used by CSPs to compute the propagation delay, clearly outlines the non-linearity of the model (between delay and skew) and the indirect impact of the temperature noise on delay estimation through the skew.

For every exchange of synchronization messages, two samples of $\theta$, namely $\check \theta_{SR}$ and $\check \theta_{RS}$ can be computed by:
\begin{eqnarray}
\check \theta_{SR}(k) & = & t_1(k)-(t_2(k)-d(k)) \label{eq:measure2} \\
\check \theta_{RS}(k) & = & t_4(k)-(t_3(k)+d(k)) \nonumber
\end{eqnarray}
The use of both equations to compute the offset, peculiar of the two-way timing message exchange mechanism, allows a better estimation of the offset.

\begin{figure}[t]
\hskip-0.5in
\centering
\includegraphics[width=6cm]{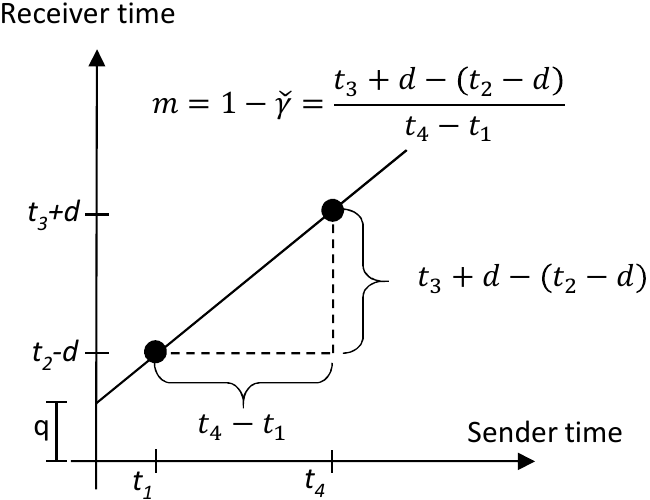}
\caption{Skew sample $\check \gamma$ from the slope $m$ of the interpolation of $t_1, t_2, t_3, t_4$.} \label{fig:GammaFromTimestamps}
\end{figure}

A sample of the skew, $\check \gamma(k)=1-m$, is derived by calculating the slope $m$ of the line interconnecting the two points $(t_1(k), t_2(k)-d(k))$ and $(t_4(k), t_3(k)+d(k))$, placed in the sender-receiver space. The slope represents the ratio between the receiver and the sender oscillation periods of the XO. More specifically, the differences $t_4$-$t_1$ and $t_3$-$t_2$+$2d$ lie on the sender and receiver time-lines, respectively, as outlined in Fig.~\ref{fig:GammaFromTimestamps}.

\begin{eqnarray}
\check \gamma(k) = 1-\frac{t_3(k)+d(k)-(t_2(k)-d(k))}{t_4(k)-t_1(k)} \label{eq:measure3}
\end{eqnarray}

An intuitive example of the application of (\ref{eq:measure1})-(\ref{eq:measure3}) is provided in \ref{sec:AppendixA}.
Operatively, the measurement equations may be simplified as follows. By assuming delay stationarity over each observation period $[k-K,..., k], k=1,2,...$, a delay estimation at time $k$, $\hat d(k)$, may be derived from the arithmetic average of $\check d(k-K), ..., \check d(k)$ and by simplifying $\gamma(k)$ in (\ref{eq:measure1}) with $\hat \gamma(k-1)$ (i.e., with the previous estimation of $\gamma$; $\hat \gamma(1)=1$). The $\hat d(k)$ quantity may be then substituted in (\ref{eq:measure2}) and (\ref{eq:measure3}). This kind of calculation is however not necessary for the proposed estimation schemes as outlined in subsection \ref{sec:delaycorrection}. The stationarity assumption on delay means that nodes mobility takes place over time horizons larger than $K$ (i.e., the size of the observation period). In case of mobility over shorter ranges, the state equation (\ref{eq:state3}) should be updated with an appropriate noise variable or with an additional function mapping $d(k-1)$ in $d(k)$.

\subsection{Measurement noise} \label{sec:measurementnoise}
The noise affecting the equations above is due to the randomness of the values of $t_1, t_2, t_3, t_4$, which derives from the chain of delays evidenced in Fig.~\ref{fig:fig_delays-crop} and defined as follows (the index $k$ is not mentioned for the sake of clarity).

By following the clock register $C_S$, at time $t_S$ (top-left of the figure), the sender schedules the sending of a synchronization packet.
After a $d_s^{node'}$ delay (corresponding to the operational times of the local operating system, the Media Access Control (MAC) and the transceiver) the packet is actually sent, and its sending time $t_1$ can be recorded by the sender node after a $d_s^{send}$ delay (corresponding to the operational time for handling the time-stamping procedure triggered by the network adapter\footnote{In order to improve precision of software timestamping, the timestamping operation is scheduled by the first instruction of the inherent Interrupt Service Routine (ISR). Moreover, the network adapter is sometimes assumed to be able to insert timestamps on-the-fly, just before the packet is transferred over the air. In this case, the third packet exploited by the protocol to deliver $t_2$ and $t_3$ to the sender is no longer required. Some other adapters acquire the timestamps in hardware, i.e., at MAC level, thus allowing the reduction of delay oscillations even more.}). The packet arrives at the receiver after $d_{SR}^{prop}$, i.e., the medium propagation delay from the sender to the receiver.

Analogous delay quantities are defined for the receiver when setting the values of $t_2$ and $t_3$ (namely, $d_R^{rec}$ and $d_R^{send}$), for the sender again when setting the value of $t_4$ (i.e., $d_S^{rec}$), and for the medium propagation delay $d_{RS}^{prop}$ for packets sent in the opposite direction. The last packet is sent from the receiver to the sender (bottom right in the figure); it contains the values of $t_2$ and $t_3$ needed for a clock correction step. It is sent after the first response packet, which triggers the $t_4$ computation at the sender.
All $t_1, t_2, t_3, t_4$ are involved in the synchronization update through (\ref{eq:measure1})-(\ref{eq:measure3}). The clock update occurs at time $t_S^{sync}$ after a delay $d_s^{node''}$ (bottom left in the figure), that includes the inherent estimation procedure and other in-node overheads.

It is important to remark that the traffic generated by other interfering nodes in the wireless network does not influence the precision and the accuracy of the timestamps.
In practice, timestamps $t_1$ and $t_3$ are recorded after the transmitting node has already started the transmission on the ether, and after a possible wait due to carrier sense.
As a consequence, $t_1$ and $t_3$ do not include the error due to the indeterminism of the access schema used by the MAC layer of the wireless communication protocol. On the receiving side, timestamps $t_2$ and $t_4$ are acquired as quickly as possible after the arrival of the synchronization message, and the only possible effect of interfering traffic and disturbs is the loss of some messages. The effect of losses for a technique based on a $1$-st order regression spline was analyzed in a real implementation in \cite{TII-sync-industrial}. Results reveal that the impact of losses on synchronization quality is negligible when $1$-st order regression spline is computed on at least $70\%$ of the expected points, and they do not worsen considerably when the number of losses further increases. It can safely assumed that the same applies to the approach based on a neural network proposed in this paper, because it uses a $1$-st order regression spline for features extraction. A more detailed analysis of this aspect, and how to further improve accuracy for the proposed technique, is left open for future research.

\section{Optimal state estimation}
In principle, the problem consists of defining the optimal filter for state estimation at each time $k$, on the basis of measurements collected up to time $k$ \cite{nostroToIM}. The optimal filter is a function\footnote{\emph{Filter}, \emph{estimation law} or \emph{estimation function} are typically used as synonyms.} $\nuv^o(\cdot)$ that maps the measurements into the estimates at each time $k$. Let $\xv(k)=\fv(\xv(k-1),\xiv(k-1))$ be the state equation in compact form from (\ref{eq:state1}), (\ref{eq:state2}) and (\ref{eq:state3}) with $\xv(k)=[\theta(k), \gamma(k), d(k)]$, $\xiv(k)$ being the vector of state noises and $\yv(k)=\gv(\xv(k),\etav(k))$ the measurement equation from (\ref{eq:measure1}), (\ref{eq:measure2}) and (\ref{eq:measure3}) with $\yv(k)=[\check \theta_{SR}(k), \check \theta_{RS}(k), \check \gamma(k), \check d(k)]$, $\etav$ being the vector of measurement noises, respectively. The optimal estimation law $\nuv_k^o(\cdot)=\nuv_k^o(\Iv_k)$ minimizes the following functional cost:
\begin{eqnarray} \label{eq:OptimalFilter}
\nuv_k^o(\Iv_k) = arg \min_ {\nuv_k(\Iv_k)} \E_{\scriptsize {\xv(k)}} \{ \hv(\xv(k)-\nuv_k(\Iv_k))  | \Iv_k \}, \forall \ \Iv_k
\end{eqnarray}
$\Iv_k$ being the information vector collecting all the measurements from the beginning $\Iv_k=[\yv(0),...,\yv(k)]$ and $\hv(\cdot)$ being a Bayesian risk function\footnote{$\hv(\zv)$ is a Bayesian risk function if the following are met: $\hv(\zv)$ is not negative, it is symmetric, i.e., $\hv(\zv)=\hv(-\zv)$ and it is not decreasing with increasing positive \zv; in the scalar case, examples are: $h(z)=z^2$ and $h(z)=|z|$. Such a risk function is used in statistical decision theory as a measure of the difference between the estimation and the true value (see, e.g., subsection 1.2.1 of \cite{Bayesian}).}. The optimal filter cannot be derived in closed-form as in the Kalman filter owing to the non-linearity of (\ref{eq:measure1}) and to the temperature noise which is not Gaussian. Here we resort to an approximating technique, as later outlined in Section \ref{sec:NN}.

\section{Splines} \label{sec:Splines}
Before addressing the approximation of the optimal filter, a basic heuristics is defined. If Fig.~\ref{fig:GammaFromTimestamps} includes the collection of $K$ sets composed of $4$ timestamps ($t_1, t_2, t_3, t_4$), the trend of the current asynchronism may be derived by interpolating the corresponding $2 K$ points in the sender-receiver space, thus deriving a heuristic estimation of $\gamma$ and $\theta$. This is the underlying idea of the splines. The following information vector is defined:
\begin{equation} \label{eq:Is}
\Iv^s_k=\left[ \varrhov^{t_2,t_1}_{k-K}, ..., \varrhov^{t_2,t_1}_{k} , \varrhov^{t_3,t_4}_{k-K}, ..., \varrhov^{t_3,t_4}_k \right]
\end{equation}
with:
\begin{eqnarray}
\varrhov_{k-j}^{t_2,t_1}=\left[t_2(k-j), t_1(k-j)\right]; \ \ \ \ j=0,...,K.\\
\varrhov_{k-j}^{t_3,t_4}=\left[t_3(k-j), t_4(k-j)\right]; \ \ \ \ j=0,...,K.
\end{eqnarray}
where $\varrhov_{h}^{t_2,t_1}$ and $\varrhov_{h}^{t_3,t_4}$ pertain the timestamps of Fig.~\ref{fig:GammaFromTimestamps} at time $h=k-j$ and $\Iv^s_k$ pertains the collection of timestamps in $[k-K, ..., k]$.\\

The $i$-th order spline (denoted by \textit{S}$_i$) is derived by interpolating, with the $i$-th order, the set of points in $\Iv^s_k$ by means of the Ordinary Least Squares method \cite{OLS}. As intuitively summarized by Fig.~\ref{fig:GammaFromTimestamps}, a skew estimation is derived from the slope of \textit{S}$_1$ and the offset estimation from putting in the spline equation the current value of time of the sender.
Quadratic and cubic splines help chase the non-linearity of the offset correction. Orders higher than 3 have been disregarded to avoid overfitting, to which the splines are more sensitive with more noise and large $K$. At the sender, the synchronization step consists of directly putting the current notion of time, $C_S$, into the spline equation, thus deriving the current notion of time at the receiver $C_R$.
Because the slope of a $1$-st order spline is an estimation of $\gamma(k)$, \textit{S}$_1$ perfectly compensate all the stationary effects on the skew (e.g., the real oscillation frequency of the XO which differs from the nominal one), and all the effects on frequency with a periodicity much greater than $\tau$, such as aging. Effects with periodicity slightly greater than $\tau$ are compensated by \textit{S}$_1$ only on average, but they can be addressed by the technique based on a neural approximation described in the next section. Examples are temperature variations or vibration with a periodicity greater than $\tau$. Periods lower than $\tau$ are typically disregarded. A remedy is the reduction of $\tau$ according to the specific working conditions.

\subsection{Delay correction} \label{sec:delaycorrection}
Let $d$ represents the knowledge of the delay, derived in any way; it may be either the propagation delay, including in-node overheads or not, or the delay estimation outlined in subsection \ref{sec:measurementequations}. This knowledge reduces the noise on the timestamps $t_2, t_3$, if we apply a correction to replace them with: $t_2-d$ and $t_3+d$. In principle, this may drive a better state estimation, but it reveals to be useless for the splines. If a sufficient number of samples is taken, the regression schemes lead to identical curves, independently to the application of the correction. Intuitively, this is due to the averaging operation operated by regression while capturing the trend of the timestamps. We empirically validated this property for all the splines considered; a formal demonstration for \textit{S}$_1$ is provided in \ref{sec:AppendixB}.

\section{Neural approximation} \label{sec:NN}
A further generalization in the direction of tracking the non-linearity and non-stationarity of the involved processes is now addressed. The focus is on the offset, thus disregarding skew and delay estimation. Similarly to the splines, the offset estimation allows the direct implementation of the synchronization step by applying, as described in the previous Section \ref{sec:Splines}, the approximating function (based either on splines or on a neural network as derived here). A standard Neural Network (NN) training is formulated as follows. A new information vector is defined:
\begin{equation} \label{eq:INN}
\Iv^{NN}_k=[\delta_{k-2K}, ..., \delta_k]
\end{equation}
$\delta_k$ being the distance between each $\varrhov_h$ in $\Iv^s_k$, as defined in (\ref{eq:Is}) and with $h=k-2K,...,k$, and the first order spline (\textit{S}$_1$) interpolating the set of points in $\Iv^s_k$. The corresponding training set is stated with $k=1,...,N$ samples of $\Iv^{NN}_k$ and the relative target $\varepsilon_k$, i.e., it is a set of $N$ tuples in the form $\langle \Iv^{NN}_k, \varepsilon_k \rangle$. The value $\varepsilon_k$ being the error of \textit{S}$_1$, evaluated for a given time $k$ in the sender timescale, in predicting the time in the receiver space. Given the \textit{S}$_1$ spline obtained for $\Iv^s_k$, $S_1^k(k)=m^k \cdot k+q^k$, the target can be computed as $\varepsilon_k=C_R(k)-S_1^k(k)$.

The optimal weights assignment $\wv^o$ is derived so that:
\begin{eqnarray} \label{eq:NeuralFilter2}
\wv^o=\arg\min_{\wv}J(\wv);J(\wv)=\sum_{k=1}^{N} [\varepsilon_k-\hat \nu_k(\Iv^{NN}_k,\wv)]^2
\end{eqnarray}
Problem (\ref{eq:NeuralFilter2}) can be solved by applying standard non-linear optimization techniques.
In particular, the proposed NN, which has been parameterized through an extensive series of experiments, has three layers: the first with $2K$ input nodes, the hidden layer with $10$ nodes characterized by hyperbolic tangent activation functions, and one linear output node in the last layer. It was trained using $8$ iterations (epochs) of the classical \textit{back-propagation} training algorithm. During the $8$ epochs, at each iteration on the training set the \textit{learning rate} was decreased linearly between $0.001$ and $0.00001$. The \textit{momentum} was set to $0.01$. In the test phase, the output of the NN is added to the estimation of the receiver time, which is performed by computing \textit{S}$_1$ on a time expressed in the sender timescale.

The sequence of approximation steps from the optimal filter to this scheme are the same referenced in \cite{nostroToIM}. Here, we stress the fact that (\ref{eq:NeuralFilter2}) is solved with respect to samples coming from non-stationary noises. This consequently leads to the adaptation of the approach to variable system conditions. In this respect, differently from \cite{temperature3, AdHoc1}, no online adaptation of the algorithm is required, and, differently from PF \cite{magazine}, the computational effort of the approach resides in computing $\wv^o$ offline.

\section{Implementation issues}

\subsection{Deployment of the neural estimator}
Three steps are crucial for the deployment of the neural estimator: acquiring the synchronization error $\varepsilon$ in (\ref{eq:NeuralFilter2}), building a database for training and applying the training procedure (i.e., solving (\ref{eq:NeuralFilter2})). The first step can be performed through either specialized devices \cite{2013-ISPCS-testPatterns} or by referring to a reference signal (e.g., a periodical actuation function \cite{TII-sync-industrial} or the deterministic expected time of reception in TDMA \cite{Ber15}). An external device is in charge as well to perform the remaining two steps.

Those steps may be hardly applied in multi-hop networks if they should be repeated for each node in the network. For this reason, we derive a method to join the replication of the steps into a single procedure in the multi-hop context (i.e., one database and one training).

\subsection{Computational cost} \label{sec:computationalcost}
One main aim of \cite{Chaudhari} is to derive a simple synchronization technique with less computational complexity than Linear Programming (LP) or other traditional approaches \cite{magazine}. This is crucial for WSNs in which energy is a scarce resource. In the LP case, for example, an optimization problem is formulated on skew and offset. The problem can be solved by traditional optimization algorithms, like the simplex one. The simplex method is efficient in practice, even though it has exponential worst-case complexity. The computational issue may become critical in the synchronization context because the number of LP constraints scales up linearly in the size of the information window collecting the history of the timestamps (see, e.g., (21) of \cite{magazine}). Despite the splines advocate the adoption of the Ordinary Least Squares method, for which similar considerations may be outlined, their computational complexity is low, in particular for \textit{S}$_1$. In the bivariate case (the regression is applied on the plane reported in Fig.~\ref{fig:GammaFromTimestamps}), each new sample contributes to the updating of the \textit{S}$_1$ parameters through trivial operations, such as the update of the sum of products on previous samples (see, e.g., \cite{AdHoc1} and \cite{AdHoc1-cite-29-book}). The \textit{NN} experiences a low computational complexity as well because it depends on the collection of \textit{S}$_1$ estimations. After building $\Iv^{NN}$ on \textit{S}$_1$, the remaining \textit{NN} operations consist of computing the \textit{NN} output through a cascade of summations and multiplications involving the neural units, typically represented by hyperbolic tangent or sigmoidal functions. The \textit{NN} used in the performance evaluation has been implemented on an Atmel ATmega328P microcontroller running at $\unit[16]{MHz}$ and tested with a \textit{NN} consisting of $15$ inputs, $10$ hyperbolic tangent hidden units and one linear output. Such a microcontroller is of common use in the WSN context. The registered mean execution time of each iteration involving both features extraction (from \textit{S}$_1$) and the computation of the output (after training) was $\unit[4.932]{ms}$ \cite{nostroToIM}. The \textit{NN} computation also scales linearly with respect to $K$. Roughly speaking, this corroborates the adoption of the \textit{NN} with $\tau\geq100$ ms, $\tau$ being the size of the synchronization time steps. An accurate calibration of $K$ and of the other \textit{NN} parameters (the number of hidden units, in particular) deserves further attention if smaller synchronization steps are required.

The computational cost of the \textit{NN} train phase is higher than the one for test. Actually, the duration of training is not a limitation because it is executed offline. Unless not differently specified, all NNs have been trained with $100000$ samples, which in real systems must be acquired at runtime through measurements. As a consequence, the time needed to acquire the train database, with $\tau=\unit[1]{s}$, is about $\unit[27]{hours}$. The offline training time, measured on a PC equipped with an Intel Core i7-3770 CPU running at $\unit[3.4]{GHz}$, with a not optimized software and with $K=60$ (i.e., $120$ inputs) is about $\unit[10]{minutes}$. This time can be reduced of at least one order of magnitude with software optimization or by exploiting GPUs \cite{2010-CUDA}.

\section{Multi-hop analysis}
In a number of operating conditions, usually in large networks, nodes communicate through intermediate devices. In the viewpoint of synchronization, the nodes are hierarchically ordered in a \textit{tree} topology \cite{magazine, AdHoc1}, where at the root of the tree lies the \textit{reference clock} (i.e., the time source of the network tree). The first layer nodes synchronize directly with the reference clock. A second layer node synchronizes with the first layer. The same applies for the subsequent layers. Each layer suffers of a worst synchronization quality as soon as the distance from the root increases.

\begin{figure}[t]
\centering
\includegraphics[width=0.85\columnwidth]{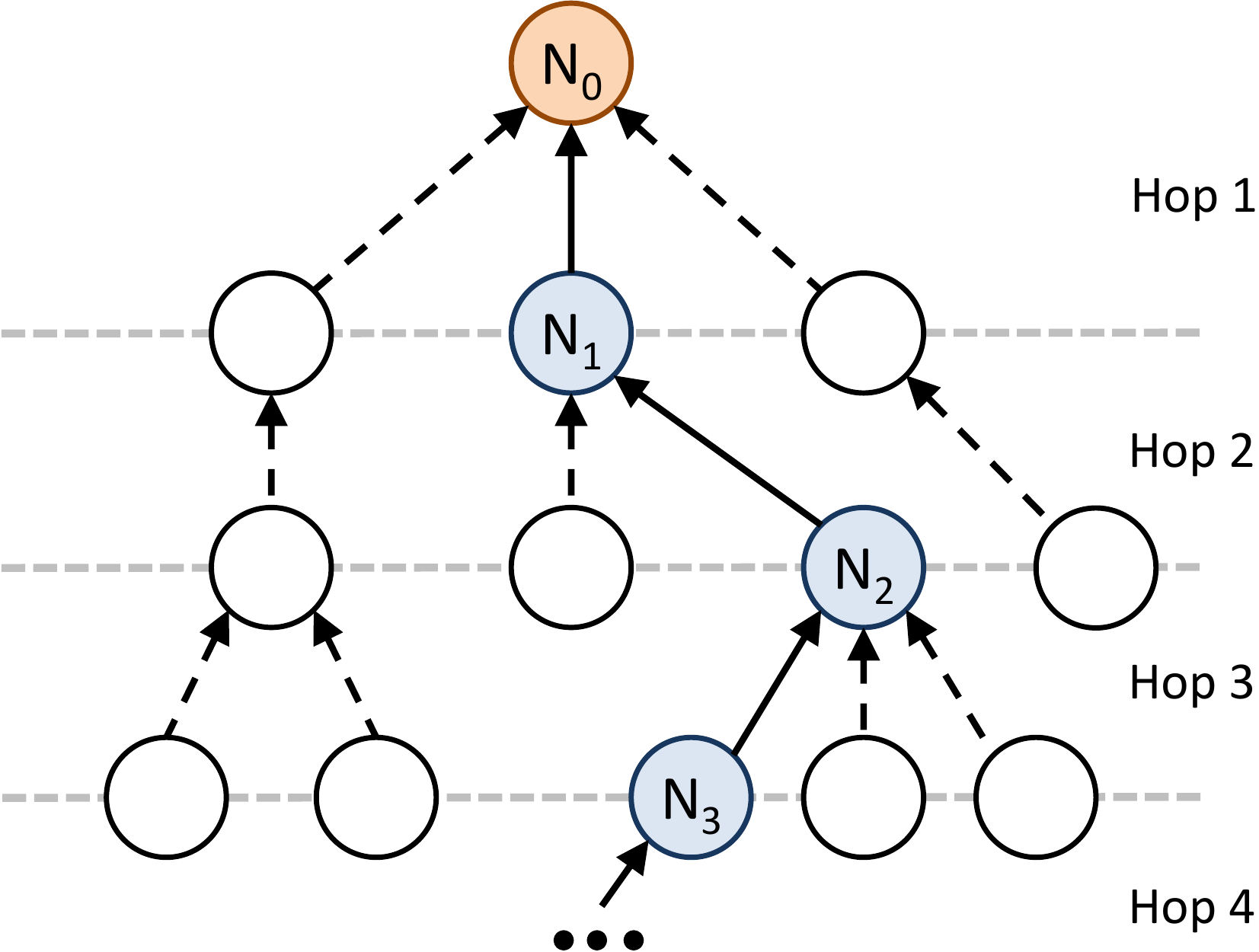}
\caption{Example of a multi-hop network.} \label{fig:multihop}
\end{figure}

Given a node, only one path exists between the node and the root of the tree. Synchronization can be separately analyzed along each path of the tree. We focus on the example reported in Fig.~\ref{fig:multihop}. The reference clock, node $N_0$, has only the \textit{master} role, i.e., it is the time source to which all other nodes must synchronize. A node of the first layer, $N_1$ has a $1$-hop distance from $N_0$, and it acts as a \textit{slave} node with respect to $N_0$ and as a master node with respect to the nodes lying in the second layer, for example $N_2$. The node $N_2$, that like $N_1$ covers both the master and slave roles, has a $2$-hop distance with $N_0$. The node $N_2$ is synchronized with $N_1$, which in turn is synchronized with $N_0$. Basically, timestamps obtained on timing messages by node $N_1$, when $N_2$ synchronizes with $N_1$, are obtained with the view of $N_1$ of the reference time held by $N_0$. In other words, $N_1$ exploits the most recent estimation of the parameters of the virtual clock to convert the timestamps obtained with its local clock to the reference time held by $N_0$. Since the virtual clock makes errors in this conversion, a node belonging to a hop level greater than 1 synchronizes its clock to an incorrect clock source. We will refer to such a kind of corruption as \textit{hop error}. The hop error increases with the number of hops and does not take place in $N_1$. Leaf nodes have only the slave role and are the ones with the biggest hop error.

Three methods based on \textit{NN}, namely \textit{NN}$_{link}$, \textit{NN}$_{gen_A}$ and \textit{NN}$_{gen_B}$, are defined in this context.

\subsection{A distinct NN for every link}
The \textit{NN}$_{link}$ model implies that the usual NN is trained link-by-link, i.e., by repeating a training phase for every communication link between two adjacent nodes of the network.
From the second hop onwards, the neural network is trained with timestamp data derived from a node which is synchronized through a NN with the node of the previous hop.
This method come out with a set of specialized NNs, able to cope with the specific experimental conditions regarding each couples of nodes. Unfortunately, this may result in a high number of database acquisitions and neural trains. Moreover, to acquire the training database for a specific hop, the NN of the previous hops must already be trained. This iterative approach leads to a complex acquisition process of the training databases and waste of time for the system setup. As a matter of fact, it is hardly applicable in real situations, except when the number of nodes is reasonably limited.

\subsection{Generalized NN: temperature compensation}
The limit of having a specific NN for every link of the network can be circumvented by training with respect to different temperature patterns. The accuracy of this kind of ``generalized'' NN is usually lower than the one of a NN trained with a database coherent with the test conditions (see, e.g., \cite{nostroToIM}).
We firstly define a NN trained over a set of temperature conditions, in-node delays, but by disregarding the hop error: \textit{NN}$_{gen_A}$. This means the training database can be easily obtained by connecting in separate, but not consecutive, experiments all the nodes with the master.

\subsection{Generalized NN: temperature and hop error compensation}
Under the \textit{NN}$_{gen_B}$ model, a single neural network is trained under an iterative (link-by-link) approach, as done in the \textit{NN}$_{link}$ case. The difference relies on the superimposition of pairs of temperature and hop error conditions, which differ from the combinations checked in the test phase.

In particular, the function $\omega_{\gamma^{T}}(t, \cdot)$ modeling the effects of temperature variation on the skew has been chosen to be different in the train and test phases. To this purpose, the temperature models evaluated in the test phase for the last and penultimate nodes drive the collection of training data at the first and second nodes, respectively, and so on. As an example, in a network with $5$ hops, $\omega^1_{\gamma^{T}}$, $\omega^2_{\gamma^{T}}$, ..., $\omega^5_{\gamma^{T}}$ are the temperature models used in the test phase for nodes $N_1$, $N_2$,...$N_5$, respectively. The temperature models associated to the same nodes (i.e., $N_1$, $N_2$,...$N_5$) in the train phase are $\omega^5_{\gamma^{T}}$, $\omega^4_{\gamma^{T}}$, ..., $\omega^1_{\gamma^{T}}$. The resulting database contains a mix of hop errors and temperatures over which the NN learns the clock correction, independently to the knowledge of its position in the tree branch. More sophisticated methods for the generation and synthesis of the training database are left open for future research.

\section{Experiments setting} \label{sec:experimentssetting}
The following simulation parameters are defined for the performance evaluation.

\emph{Observation horizon}. The value $K$ is the number of sets of timestamps $\{t_1, t_2, t_3, t_4\}$ used to decide the synchronization correction. $K$ is used as a variable parameter in the results (it appears in the x-axis of all the figures) in order to emphasize how the techniques may be sensitive to it. As signal processing techniques estimate stationary states, they achieve optimal performance only for a strict range of $K$ in non-stationary conditions, such as in the presence of temperature variations.

\emph{Synchronization period}. In the proposed simulation model, the sender node starts a synchronization step in a cyclic fashion, and each exchange is triggered with a period $\tau$ fixed to $\unit[1.0]{s}$. Periods greater than $\unit[1.0]{s}$ could be useful in contexts where power consumption is a main target, such as in WSNs. The effects of the parameter $\tau$ on synchronization quality has been analyzed in a specific experimental campaign in subsection \ref{subsec:tau}.

\emph{Performance metric}. The $99.9$ percentile of the synchronization errors is the performance metric. It is denoted by $p99.9$ and represents the $99.9$ percentile of the absolute difference between the reference time (i.e., the time at the receiver node) and the estimated time by the sender at the end of each timestamps exchange, as outlined in Fig.~\ref{fig:fig_delays-crop}. All the performed simulations (under a fixed $K$) contain $100000$ samples of the dynamic system (\ref{eq:state1})-(\ref{eq:state3}) whose evolution follow the temperature and delay models presented below.

The average error is disregarded because it represents only the systematic part of the error (i.e., the \emph{accuracy}), but it does not provide any information about the \emph{precision} of synchronization \cite{ISO-metrology}. Two nodes may be synchronized, on average, while still experiencing large synchronization errors; $p99.9$ helps capture a threshold limit of those errors (in the 99.9\% of the cases).

\emph{Temperature}. The $\omega_{\theta}$ and $\omega_{\gamma}$ components of equations (\ref{eq:state1}) and (\ref{eq:state2}) have variances $\sigma^2_{\theta}=\unit[10^{-17}]{s^2}$ and $\sigma^2_{\gamma}=\unit[10^{-19}]{}$, respectively. As far as the temperature is considered, two models have been taken into account. The $\omega^{high}_{\gamma^{T}}(t,\cdot)$ model represents a fast temperature variation of the XO ($t=\unit[600]{s}$) in a wide range of temperatures ($T^{Low}_E=\unit[-10]{^{\circ}C}$ and $T^{High}_E=\unit[40]{^{\circ}C}$); while the $\omega^{norm}_{\gamma^{T}}(t,\cdot)$ model is characterized by slower temperature variation ($t=\unit[1200]{s}$) than $\omega^{high}_{\gamma^{T}}(t,\cdot)$, in a narrower range of temperature with extremes $T^{Low}_E=\unit[10]{^{\circ}C}$ and $T^{High}_E=\unit[35]{^{\circ}C}$. For both temperature models $tc=\unit[60]{s}$. The models are applicable to mobile nodes in reality. An example may be an automatic forklift that enters into and exits from an industrial oven. Outdoor exposure is applicable as well \cite{Boa10, Xu16}.

\emph{Delay}. We consider two effects on delay: the propagation over the channel, $d_{SR}^{prop}$ and $d_{RS}^{prop}$, and the \emph{in-node} delays ($d_S^{send}$, $d_S^{rec}$, $d_R^{send}$ and $d_R^{rec}$), as defined in the following. The values $d_{SR}^{prop}=\unit[150]{ns}$ and $d_{RS}^{prop}=\unit[150]{ns}$ have been used in all the simulated scenarios. The value of $\unit[150]{ns}$ was chosen because it represents a reasonable distance of about $50$ meters between wireless nodes. In fact, an electromagnetic signal which a speed of $\sim \unit[3 \cdot 10^8]{m/s}$ takes $\unit[166]{ns}$ to cover $\unit[50]{m}$. The propagation delay is also considered as exponentially distributed in the last experiments of subsection \ref{sub:exp3}. The other delays of Fig.~\ref{fig:fig_delays-crop}, unless otherwise specified, have been set to $0$ (i.e., $d_S^{node'}=d_S^{node''}=d_R^{node}=0$).

\emph{Comparison with} \cite{Chaudhari}. As a performance comparison, in the method summarized by the \textit{CH} acronym, formulas (10) and (11) of \cite{Chaudhari} have been chosen for the estimation of the skew and offset, respectively. In the proposed experimental setups, they provide the best results with respect to the other variants presented in the same paper.

\subsection{In-node delays}\label{sub:innode_delays}
\begin{table}
  \caption{Means and standard deviations of in-node latencies}
  \label{tab:simParam}
  \begin{center}
    \small
    \begin{tabular}{l|l||c|c|c|c}
      \multicolumn{2}{r||}{Scenario}  & \textit{$sw^{WiFi}$} & \textit{$hw^{WiFi}$} & \textit{$sw^{WSN}$} & \textit{$hw^{WSN}$} \\
      \multicolumn{2}{r||}{Type} & PC-WiFi & PC-WiFi & Mica2 & TelosB \\
      \multicolumn{2}{l||}{Latency}  & ($\unit[]{\mu s}$) & ($\unit[]{\mu s}$) & ($\unit[]{\mu s}$) & ($\unit[]{\mu s}$) \\
      \hline \hline
      $d_S^{send}$ & $\mu_S^{send}$   & 5.4 & 1.31 & 259.057 & 0.408 \\
                  & $\sigma_S^{send}$ & 0.310 & 0.046  & 1.291 & 0.0157 \\
      \hline
      $d_S^{rec}$ & $\mu_S^{rec}$    & 7.23 & 8.9 & 346.849 & 2.769 \\
                  & $\sigma_S^{rec}$ & 0.580 & 0.110  & 2.415 & 0.0374 \\
      \hline \hline
      $d_R^{send}$ & $\mu_R^{send}$    & $n_{\mu}\cdot$5.4 & $n_{\mu}\cdot$1.31  & $n_{\mu}\cdot$259.057 & $n_{\mu}\cdot$0.408 \\
                  & $\sigma_R^{send}$ & $n_{\sigma}\cdot$0.310 & $n_{\sigma}\cdot$0.046  & $n_{\sigma}\cdot$1.291 & $n_{\sigma}\cdot$0.0157  \\
      \hline
      $d_R^{rec}$ & $\mu_R^{rec}$    & $n_{\mu}\cdot$7.23 & $n_{\mu}\cdot$8.9 & $n_{\mu}\cdot$346.849 & $n_{\mu}\cdot$2.769 \\
                  & $\sigma_R^{rec}$ & $n_{\sigma}\cdot$0.580 & $n_{\sigma}\cdot$0.110 & $n_{\sigma}\cdot$2.415 & $n_{\sigma}\cdot$0.0374 \\
      \hline \hline
    \end{tabular}
  \end{center}
\end{table}
In-node delays represent the latency between the sending or reception times of the synchronization packet and when the timestamp is actually obtained. In absence of nodes' mobility or other structural changes of the environment, which may cause variations of the fading affecting the channel, the propagation delay is deterministic and the prevailing effect on synchronization is due to in-node delays.

The setting of the noise on in-node delay is now detailed. In-node delays reported in Fig.~\ref{fig:fig_delays-crop} are more concisely represented with: $d_{\rho}^{send}$ and $d_{\rho}^{rec}$, where ${\rho}$ represents the node role, ${\rho}=S$ for the \textit{sender} node and ${\rho}=R$ for the \textit{receiver} node, respectively.

In \cite{TII_jitter}, in-node delays have been analyzed for IEEE 802.11 WiFi devices and results are reported for hardware and software timestamps. The resulting delay model is summarized in the first two columns of Table \ref{tab:simParam}. The two latencies $d_{\rho}^{send}$ and $d_{\rho}^{rec}$ are not symmetric, and sender and receiver have very different in-node delays values in terms of both mean and standard deviation (for instance $sw^{WiFi}$ and $hw^{WiFi}$ in Table \ref{tab:simParam}).

The presented setting for software time-stamping (condition $sw^{WiFi}$ reported as first column of Table \ref{tab:simParam}), is referred to a system with low interfering loads (i.e., CPUs often in the \textit{IDLE} state, low interrupts rate, etc.).
With hardware timestamps, $hw^{WiFi}$ condition, the standard deviations $\sigma_{\rho}^{send}$ and $\sigma_{\rho}^{rec}$ are quite small.

The distribution of $d_{\rho}^{send}$ and $d_{\rho}^{rec}$ \cite{TII_jitter} may have various shapes depending on the nodes hardware, operating system and internal load. A good approximation is however the normal distribution: $d_{\rho}^{send}=\mathcal{N}(\mu_{\rho}^{send}, \sigma_{\rho}^{2\ send})$ and $d_{\rho}^{rec}=\mathcal{N}(\mu_{\rho}^{rec}, \sigma_{\rho}^{2\ rec})$.
The multipliers $n_{\mu}$ and $n_{\sigma}$ in Table \ref{tab:simParam} are used in subsection \ref{sub:exp2} in order to set variable asymmetry conditions on the delays $d_{\rho}^{send}$ and $d_{\rho}^{rec}$ of the receiver node. The quantities $d_{\rho, i}^{send}-\mu_{\rho}^{send}$ and $d_{\rho, i}^{rec}-\mu_{\rho}^{rec}$ are usually known as \emph{jitter}.

As far as WSNs are considered, some papers \cite{AdHoc1, tesi_jitter, FTSP} have experimentally evaluated and analyzed the distributions of $d_{SR}^{path}$ and $d_{RS}^{path}$, where:
\begin{eqnarray}
d_{SR}^{path} & = & d_{SR}^{prop}+d_R^{rec}-d_S^{send} \nonumber \\
d_{RS}^{path} & = & d_{RS}^{prop}+d_S^{rec}-d_R^{send} \nonumber
\end{eqnarray}
represent the measured path delays between sender and receiver, and vice versa. At the best of authors knowledge, a separate analysis for in-node delays is not currently available. In \cite{tesi_jitter}, the transmission latency of a message between WSN nodes of different type have been evaluated. In particular, for a Mica2 WSN node the reported delay is $\unit[605.906]{\mu s}$, with a standard deviation equal to $\unit[2.738]{\mu s}$.

In order to have a coherent model for WSN, we split the values reported in \cite{tesi_jitter} between the $d_{\rho}^{send}$ and $d_{\rho}^{rec}$ contributions, with the constraint of maintaining the ratios $\frac{\mu_{\rho}^{send}}{\mu_{\rho}^{rec}}$ and $\frac{\sigma_{\rho}^{2\ send}}{\sigma_{\rho}^{2\ rec}}$ experienced for $sw^{WiFi}$. The previously described process was used to obtain the $sw^{WSN}$ condition (third column of Table \ref{tab:simParam}).

\begin{figure*}[t]
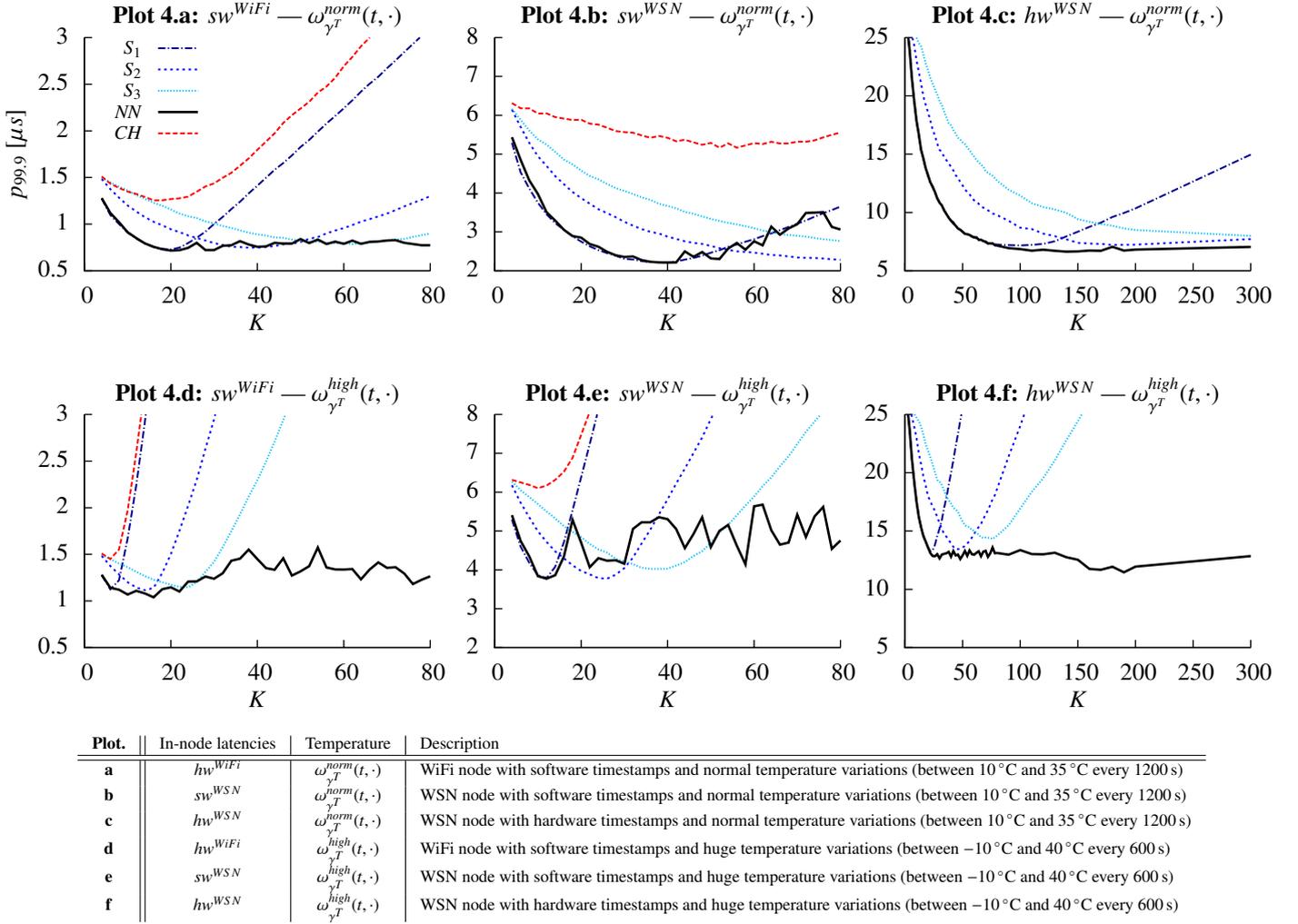

\centering
\include{fig4}
\vspace{-0.3cm}
  \begin{center}
    \scriptsize
    \begin{tabular}{c||c|c|l}
\textbf{Plot.} & In-node latencies & Temperature & Description \\
\hline\hline
\textbf{a}  & $hw^{WiFi}$ & $\omega^{norm}_{\gamma^{T}}(t,\cdot)$ & WiFi node with software timestamps and normal temperature variations (between $\unit[10]{^{\circ}C}$ and $\unit[35]{^{\circ}C}$ every $\unit[1200]{s}$)\\
\textbf{b}  & $sw^{WSN}$ & $\omega^{norm}_{\gamma^{T}}(t,\cdot)$ & WSN node with software timestamps and normal temperature variations (between $\unit[10]{^{\circ}C}$ and $\unit[35]{^{\circ}C}$ every $\unit[1200]{s}$)\\
\textbf{c}  & $hw^{WSN}$ & $\omega^{norm}_{\gamma^{T}}(t,\cdot)$ & WSN node with hardware timestamps and normal temperature variations (between $\unit[10]{^{\circ}C}$ and $\unit[35]{^{\circ}C}$ every $\unit[1200]{s}$)\\
\textbf{d}  & $hw^{WiFi}$ & $\omega^{high}_{\gamma^{T}}(t,\cdot)$ & WiFi node with software timestamps and huge temperature variations (between $\unit[-10]{^{\circ}C}$ and $\unit[40]{^{\circ}C}$ every $\unit[600]{s}$)\\
\textbf{e}  & $sw^{WSN}$ & $\omega^{high}_{\gamma^{T}}(t,\cdot)$ & WSN node with software timestamps and huge temperature variations (between $\unit[-10]{^{\circ}C}$ and $\unit[40]{^{\circ}C}$ every $\unit[600]{s}$)\\
\textbf{f}  & $hw^{WSN}$ & $\omega^{high}_{\gamma^{T}}(t,\cdot)$ & WSN node with hardware timestamps and huge temperature variations (between $\unit[-10]{^{\circ}C}$ and $\unit[40]{^{\circ}C}$ every $\unit[600]{s}$)\\
\end{tabular}
\end{center}

\caption{99.9-percentile of the synchronization error, for \textit{S}$_1$, \textit{S}$_2$, \textit{S}$_3$, \textit{NN} and \textit{CH} ($\omega^{high}_{\gamma^{T}}(t,\cdot)$ and $\omega^{norm}_{\gamma^{T}}(t,\cdot)$ temperature models; $sw^{WiFi}$ and $sw^{WSN}$ in-node latencies; $n_{\mu}=1$ and $n_{\sigma}=1$).} \label{fig:exp_comparison}
\end{figure*}

The same splitting procedure, between the sending and the receiving components of the delay, has been used for the $hw^{WSN}$ condition. In this case, data about delays have been obtained from \cite{AdHoc1}, and the constraint regarding the two ratios are coherently referred to a hardware condition, i.e., $hw^{WiFi}$. In \cite{AdHoc1}, the experimental setup is composed of TelosB WSN motes, and the measured delay of a transmission between two nodes of this type --- including in-node delays but excluding the timestamps quantization errors --- has a mean value and a standard deviation of $\unit[3.177]{\mu s}$ and $\unit[40.56]{ns}$, respectively. An important distinguish characteristic of the $hw^{WSN}$ scenario is that the timestamps resolution is $\sim\unit[31]{\mu s}$, because nodes local clock has a frequency of $\unit[32768]{Hz}$. As a result, even if the precision of the timestamp is very high (i.e., the standard deviation is below $\unit[50]{ns}$), the use of a great number of timestamps is mandatory to mitigate the effects related to their low resolution.

The use of more than one timestamp to drive estimation (i.e., $K>1$) helps mitigate in-node jitter or quantization errors for all the conditions, with the only exception of $hw^{WiFi}$, whose timestamps are characterized by a higher precision.
Under $hw^{WiFi}$ conditions, small values of $K$, e.g., $K=2$, are sufficient for synchronization with \textit{S}$_1$. Under peculiar circumstances, e.g., with high values of $\tau$, or communication losses between sender and receiver, further adjustments may be needed; this has been left open for future research.

\section{Performance evaluation} \label{sec:performance}

\subsection{Simulation environment} \label{sec:sim_environment}
Typical network simulators are designed to model adequately communication protocols, and they can easily scale to large size networks composed of a high number of nodes. Examples of popular network simulators are the open source ns-2, ns-3 and Omnet++, or the commercial solution OPNET modeler. Unfortunately, for clock synchronization, and in particular for those aspects mostly taken into account here (i.e., effects of temperature and in-node latencies on synchronization quality), available network simulators are not adequate. Firstly, they cannot currently model XOs and in-node latencies. Both aspects can be in theory included by modifying simulator models, but this requires a good knowledge of the simulator, a lot of effort, and there is no guarantee that changes will be compatible with newer versions of the simulator. Secondly, network simulators are not designed to model more than one timescale. In synchronization, each node has a different view of the time, and its dependence with respect to the timescale of the simulator is complex (see, e.g., subsections \ref{sub:state_equations} and \ref{sub:state_noise}).
Finally, to model the schema proposed in Fig.~\ref{fig:fig_delays-crop}, if each message exchange starts and end in $[k, k+1], \forall k$, a discrete event simulator is not needed. In fact, timestamps acquired by nodes in the exchange started at time $k+1$ do not depend on that acquired in the exchange $k$. This reasonable assumption reduces considerably the complexity of the simulator.

For all these reasons, an ``ad-hoc'' simulation environment has been specifically developed to model the system as described in the previous sections. The simulation software was programmed in \texttt{python}, and it was executed in parallel (a process for every value of $K$) in a High-Performance Computing (HPC) cluster consisting of $544$ cores placed in $17$ computational nodes, and with a total amount of RAM equal to $\unit[2.2]{TB}$. The simulation process is subdivided in two phases. In the first phase, the simulator makes use of state equations (including noises, environmental temperature variation and XO models) described in Section \ref{sec:problem}, to obtain for each time instant the four timestamps exploited by the clock correction algorithm (i.e, $t_1$, $t_2$, $t_3$ and $t_4$), and the correct target time. In the second phase, data obtained in the first step are exploited to compare clock correction algorithms. Splitting the simulation into two steps offers a big advantage in terms of execution speed, because data have not to be reproduced each time the performance of a clock correction algorithm has to be tested, and it ensures a fair comparison between algorithms \cite{2015-etfa-syncDB}, i.e., all are applied to the same data set.

\subsection{Varying nodes type and environmental conditions}
In the first set of experiments (Fig.~\ref{fig:exp_comparison}), we evaluate the performance of all the techniques and temperature models proposed, with the exception of $hw^{WiFi}$. For the $hw^{WSN}$ condition, the delays $d_S^{node'}$ and $d_R^{node}$ have been distributed uniformly between $0$ and $\unit[31]{\mu s}$. This setting makes it possible to put out of phase the sending times of the two exchanged packets. This procedure also removes possible correlations between the timestamps obtained in subsequent synchronization steps.

It is clear from the figure that \textit{CH} is never optimal. The splines, especially \textit{S}$_1$, guarantee the optimal performance only for short ranges of $K$. The optimal setting (i.e., minimum $p99.9$) of $K$ is denoted with $K^*$. Under high temperature oscillations (i.e., $\omega^{high}_{\gamma^{T}}(t,\cdot)$), both the splines and \textit{CH} have a significant performance degradation as soon as $K$ slightly differs from $K^*$. The value $K^*$ is not constant and it depends on the spline used, on the temperature model and on the in-node delays. As a consequence, the estimation of $K^*$ is hardly possible. Conversely, the \textit{NN} guarantees optimal performance for larger ranges of $K$. Only the most critical condition of Plot~\ref{fig:exp_comparison}.e (software timestamping in WSN and high temperature variations) leads to larger performance oscillations in the \textit{NN}. The method based on \textit{NN} provides lower synchronization errors also in the case of hardware timestamps with $\unit[31]{\mu s}$ resolution ($hw^{WSN}$ condition), outperforming all the other methods.

\begin{figure}[t]
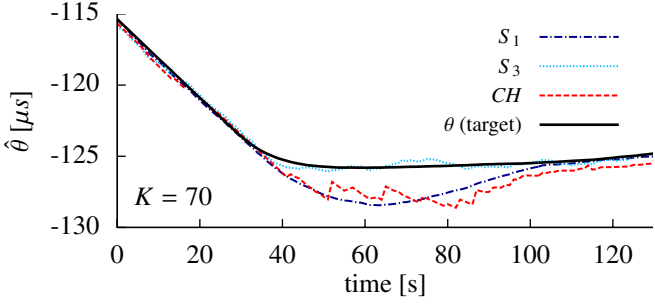

\centering
\include{fig5}
\caption{Estimation of the offset ($\hat \theta$), with different time-varing techniques, in proximity of a trend inversion at time $\unit[40]{s}$ ($K=70$).}
\label{fig:track}
\end{figure}

In Fig.~\ref{fig:track}, the \textit{S}$_1$, \textit{S}$_3$ and \textit{CH} techniques have been analyzed in the proximity of a trend inversion (at time $\unit[40]{s}$, the temperature reaches the minimum, $\sim T^{Low}_E$, and starts to increase again). Fig.~\ref{fig:track} helps highlight the impact of tracking the variability of the target offset, which is slowly approximated by \textit{S}$_1$ and \textit{CH}. In this case, with $K=70$, the best tracking is obtained by \textit{S}$_3$ and \textit{NN} (the \textit{NN} is not reported for the sake of clarity). Under small values of $K$, however, the use of high order splines is not convenient, because the contribution of the measurement noise is predominant with respect to the temperature effect. In virtue of the small sensitivity on $K$ variations, \textit{NN} is not affected by this problem.

\subsection{Synchronization period $\tau$}\label{subsec:tau}

\begin{figure}[t]
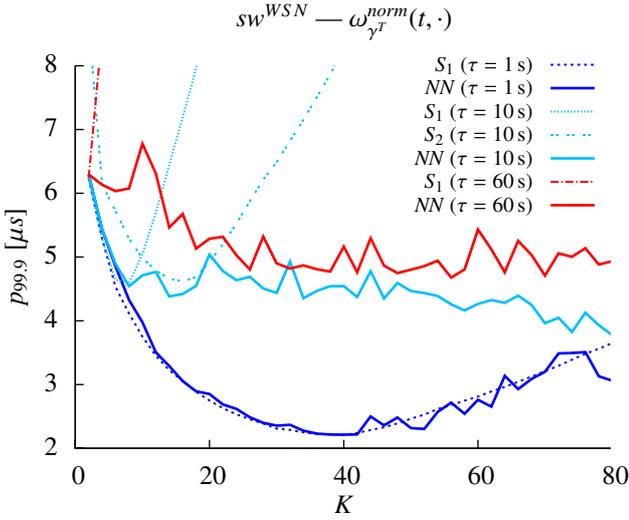

\centering
\include{fig6}
\caption{99.9-percentile of the synchronization error, for \textit{S}$_1$ and \textit{NN} with different values of the synchronization period ($\tau = 1, 10, \unit[60]{s}$). For $\tau=\unit[10]{s}$ also the results of the \textit{S}$_2$ technique has been reported. Experimental condition: $sw^{WSN}$ --- $\omega^{norm}_{\gamma^{T}}(t,\cdot)$, i.e., WSN node with software timestamps and normal temperature variations (between $\unit[10]{^{\circ}C}$ and $\unit[35]{^{\circ}C}$ every $\unit[1200]{s}$).}
\label{fig:exp4}
\end{figure}

\begin{figure}[t]
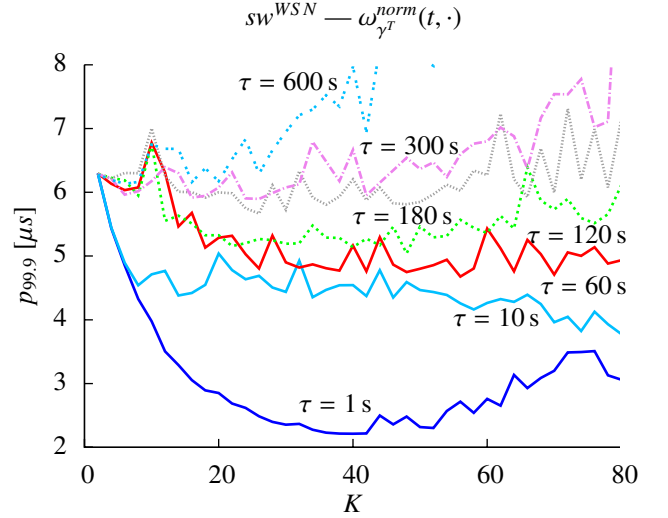

\centering
\include{fig7}
\caption{99.9-percentile of the synchronization error, for \textit{NN} with different values of the synchronization period ($\tau = 1, 10, 60, 120, 180, 300, \unit[600]{s}$). Experimental condition: $sw^{WSN}$ --- $\omega^{norm}_{\gamma^{T}}(t,\cdot)$, i.e., WSN node with software timestamps and normal temperature variations (between $\unit[10]{^{\circ}C}$ and $\unit[35]{^{\circ}C}$ every $\unit[1200]{s}$).}
\label{fig:exp4b}
\end{figure}

In many WSN applications, the interval between two adjacent synchronization steps, $\tau$, is usually increased as much as possible to improve battery duration. Unfortunately, an increasing $\tau$ (leaving $K$ unchanged) decreases the ability to track temperature variations. To compensate large $\tau$, a reduction of $K$ may be applied, but this would lead to worst in-node delays compensations. The estimation performance thus plays an important role in this perspective as well.

A number of experiments with variable $\tau$ have been carried out for the $sw^{WSN}$ scenario, the $\omega^{norm}_{\gamma^{T}}(t,\cdot)$ temperature model, and with different values of $\tau$, namely $\tau=\unit[1]{s}, \unit[10]{s}, \unit[60]{s}$.

Results reported in Fig.~\ref{fig:exp4} show how synchronization error is directly related to $\tau$. For instance, increasing of one order of magnitude the value $\tau$, from $\unit[1]{s}$ to $\unit[10]{s}$, the error in correspondence to $K^{*}$ passes from $\unit[2.211]{\mu s}$ to $\unit[5.590]{\mu s}$ for \textit{S}$_1$ and from $\unit[2.212]{\mu s}$ to $\unit[4.162]{\mu s}$ for \textit{NN}. After a further increase to $\tau=\unit[60]{s}$, the synchronization error worsens and it reaches in the point $K^*$ the minimum errors of $\unit[6.245]{\mu s}$ and $\unit[4.678]{\mu s}$ for \textit{S}$_1$ and \textit{NN}, respectively.

For $\tau=\unit[10]{s}$, also the statistics related to \textit{S}$_2$ has been reported in the plot. The minimum achieved error is $\unit[4.621]{\mu s}$ with $K^*=16$. The errors of \textit{S}$_1$ and \textit{S}$_2$ are higher than the one of the \textit{NN} method, for every value of $K$ and $\tau$.

With values of $\tau$ greater than a given threshold (i.e., $\tau \ge \unit[180]{s}$ as in Fig.~\ref{fig:exp4b}), synchronization quality worsen because \textit{NN} method is no longer able to estimate correctly the temperature variations. In other words, the error achieved by \textit{NN} in compensating temperature variations is bigger than the error due to in-node delay.

\subsection{Delay asymmetry}\label{sub:exp2}
In Fig.~\ref{fig:asym}, the effect of latency asymmetries on synchronization accuracy are analyzed. Tests have been performed using $sw^{WiFi}$ and $\omega^{norm}_{\gamma^{T}}(t,\cdot)$ as temperature model. Plot~\ref{fig:asym}.a represents a perfect symmetry where both sender and receiver experience the same latency distributions (i.e., $n_{\mu}=1$ and $n_{\sigma}=1$ in Table \ref{tab:simParam}). Results for this configuration are equal to those discussed in the previous subsection.

\begin{figure*}[t]
\centering
\include{fig8}
\vspace{-0.3cm}
  \begin{center}
    \footnotesize
    \begin{tabular}{c||c|c|l}
\textbf{Plot.} & $n_\mu$ & $n_\sigma$ & Description \\
\hline\hline
\textbf{a}  & $1$ & $1$ & Perfect symmetry between sender and receiver in-node latencies \\
\textbf{b}  & $1$ & $1.5$ & Asymmetry on in-nodes latencies precision was obtained by increasing the width of the receiver node in-node latencies \\
\textbf{c}  & $1.5$ & $1$ & Asymmetry on in-nodes latencies accuracy was obtaining by adding a systematic error in the receiver node in-node latencies \\
\textbf{d}  & $1.5$ (test) & $1$ & Same test configuration regarding asymmetry of Plot c. Three different train configurations: \\
 & & & \textit{NN}$_{n_{\mu}=1}$ equal to Plot a; \textit{NN}$_{n_{\mu}=1}$ equal to Plot c; \textit{NN}$_{gen}$ obtained mixing different $n_\mu$ values \\
\end{tabular}
\end{center}

\caption{The effect of in-node asymmetries on the 99.9-percentile of the synchronization error, for \textit{S}$_1$, \textit{S}$_2$, \textit{S}$_3$, \textit{NN} and \textit{CH} ($\omega^{norm}_{\gamma^{T}}(t,\cdot)$ temperature models; $sw^{WiFi}$ in-node latencies).} \label{fig:asym}
\end{figure*}

In the second experiment (Plot~\ref{fig:asym}.b), an asymmetry on the width of the gaussian distributions that model the latencies has been introduced by setting $n_{\sigma}=1.5$. As expected, the synchronization quality worsen for all the analyzed techniques, because the precision of the timestamps at the receiver has been reduced. The technique mostly affected by the asymmetry is \textit{CH}.

In the third experiment (Plot~\ref{fig:asym}.c), the asymmetry has been obtained adding a systematic error on timestamps acquisition, by multiplying only the mean value of the gaussian distributions of the receiver node: $n_{\mu}=1.5$ and $n_{\sigma}=1$. Only the \textit{NN} technique compensates the asymmetry, while other techniques suffer of a degradation on accuracy of about $\unit[3.2]{\mu s}$.

Experimental results reflect directly the systematic accuracy error introduced by the asymmetry. Since the delay sampling of equation (\ref{eq:measure1}) does not hold under asymmetry (the multiplication by $\frac{1}{2}$ assumes path symmetry), asymmetry may cause a systematic error, whose correction requires an accurate calibration (as mentioned, for example, in \cite{TII_jitter}). The calibration may be driven by a-priori calculations as shown in \ref{sec:AppendixC}. Since a-priori calculations may be hardly applied in practice, the calibration may derive from the direct measurement of the synchronization error.
The calibration provided by the \textit{NN}, together with the compensation of other non-stationary effects (e.g., temperature), greatly simplifies the synchronization process.

The fourth experiment (Plot~\ref{fig:asym}.d) analyzes the robustness of \textit{NN} with respect to the asymmetry of the channel. All tests are performed using $n_{\mu}=1.5$. Results, reported as \textit{NN}$_{n_{\mu}=1}$ and \textit{NN}$_{n_{\mu}=1.5}$, denote \textit{NN} models trained with $n_{\mu}=1$ and $n_{\mu}=1.5$, respectively. For \textit{NN}$_{gen}$, the \textit{NN} model has been trained with five data sets composed of $25000$ samples that differ on the values of $n_{\mu}$ ($0.25$, $0.75$, $1$, $1.25$, $1.75$, respectively). As expected, \textit{NN} cannot generalize the channel asymmetry as in the case of temperature \cite{nostroToIM}, because formula (\ref{eq:measure1}) supposes $d_{SR}^{path}=d_{RS}^{path}$. Basically, when the \textit{NN} is trained by using data sets with different channel asymmetries, it reaches a minimum error only for one of the possible asymmetries. \textit{NN}$_{n_{\mu}=1.5}$ has the same systematic error in all the cases. This error can be derived from the calculations presented in \ref{sec:AppendixC}. When the test and training conditions are consistent from the viewpoint of channel asymmetry, \textit{NN} compensates the error.

\subsection{Exponentially distributed delay}\label{sub:exp3}
\begin{figure*}[t]
\centering
\include{fig9}
\vspace{-0.3cm}
  \begin{center}
    \footnotesize
    \begin{tabular}{c||c|c|l}
\textbf{Plot.} & Information & Delay & Description \\
               & vector      & model &             \\
\hline\hline
\textbf{a}  & \textit{NN}$^{S_1}$ & $exp(\frac{1}{\unit[1]{\mu s}})$ & Information vector based on $S_1$, and only exponentially distributed delay \\
\textbf{b}  & \textit{NN}$^{S_1^{CH}}$ & $exp(\frac{1}{\unit[1]{\mu s}})$ & Information vector based on $S_1^{CH}$, and only exponentially distributed delay \\
\textbf{c}  & \textit{NN}$^{S_1^{CH}}$ & $sw^{WiFi}$ & Information vector based on $S_1^{CH}$, and only Gaussian in-node delay \\
\textbf{d}  & \textit{NN}$^{S_1^{CH}}$ & $sw^{WiFi}+exp(\frac{1}{\unit[1]{\mu s}})$ & Information vector based on $S_1^{CH}$, and both Gaussian in-node delay and exponentially distributed delay \\
\end{tabular}
\end{center}

\caption{99.9-percentile of the synchronization error, with different delays distributions ($sw^{WiFi}$ and $exp(\frac{1}{\unit[1]{\mu s}})$), and evaluation of the \textit{NN}$^{S_1^{CH}}$ method.} \label{fig:exp}
\end{figure*}
For the sake of completeness, we consider also exponentially distributed delays as in \cite{Chaudhari, ICCSvedesi}, which may characterize peculiar scenarios \cite{ICCSvedesi}. In the first two experiments of Fig.~\ref{fig:exp}, no in-node delays are considered, while $d_{SR}^{prop}$ and $d_{RS}^{prop}$ are exponentially distributed with mean $\unit[1]{\mu s}$. This condition has been denoted as $exp(\frac{1}{\unit[1]{\mu s}})$ in Fig.~\ref{fig:exp}.

In the first Plot~\ref{fig:exp}.a, \textit{CH} achieves the best performance. The \textit{NN} tries to follow the performance of the best spline, which changes from \textit{S}$_1$ to \textit{S}$_3$ with increasing $K$.

In the second Plot~\ref{fig:exp}.b, the \textit{NN} information vector is collected on the basis of a first-order regression scheme, whose slope and y-intercept coefficients are derived from \textit{CH} (denoted with \textit{S}$_1^{CH}$). Results regarding this new information vector have been reported as \textit{NN}$^{S_1^{CH}}$, while \textit{NN}$^{S_1}$ identifies those derived from \textit{S}$_1$. More specifically, the \textit{S}$_1^{CH}$ is obtained from computing the slope and the y-intercept of \textit{S}$_1$ through the skew and offset estimated by equations (10) and (11) in \cite{Chaudhari}, respectively. As outlined in Fig.~\ref{fig:GammaFromTimestamps}, skew and \textit{S}$_1$ slope are strictly related. With exponential distributed delay, the \textit{NN}$^{S_1^{CH}}$ method guarantees the best performance with a larger set of $K$. The quality of synchronization is always better, regardless the value of $K$.

The new information vector based on \textit{S}$_1^{CH}$ is analyzed in the third experiment (Plot~\ref{fig:exp}.c) with Gaussian in-node delays ($sw^{WiFi}$) and with $d_{SR}^{prop}=d_{RS}^{prop}=0$. Surprisingly, the synchronization accuracy of \textit{NN}$^{S_1^{CH}}$ is comparable with the one of \textit{NN}$^{S_1}$ (Plot~\ref{fig:exp_comparison}.a). This means that the information vector based on \textit{S}$_1^{CH}$ generalizes the \textit{NN} technique to both in-node Gaussian and exponential delays.

As a further verification of this property, in the last Plot~\ref{fig:exp}.d, both Gaussian $sw^{WiFi}$ and exponential $exp(\frac{1}{\unit[1]{\mu s}})$ delays have been activated in the simulator. Once again, the results highlight the capability of \textit{NN}$^{S_1^{CH}}$ to generalize to different delays distributions.

\subsection{Multi-hop scenario}
The $sw^{WiFi}$ and $sw^{WSN}$ conditions have been analyzed as they lead to the lower synchronization quality. Five hops are considered as in \cite{AdHoc1}. Table \ref{tab:multi-hop1} lists the parameters of the function $\omega_{\gamma^{T}}(t,\cdot)$ for the five hops. The parameters assigned to the different hops involve all the parameter of the function $\omega_{\gamma^{T}}(t,\cdot)$. In the first three hops, changes regard temperature boundaries (i.e., $T^{Low}_E$ and $T^{High}_E$); in the fourth hop, variations on both temperature boundaries and thermal time constant have been considered; the last hop takes into account a different periodicity (e.g., the node is placed in another operating environment, which is characterized by a different temperature variation period).

\begin{table}
  \caption{Environment characterization (i.e., parameters of the function $\omega_{\gamma^{T}}(t,\cdot)=\omega_{\gamma^{T}}(t,tc,p,T^{High}_E,T^{Low}_E)$) for every hop of the multi-hop scenario.}
  \label{tab:multi-hop1}
  \begin{center}
    \begin{tabular}{c||cccc}
hop & $T^{Low}_E$ & $T^{High}_E$ & $tc$ & $p$ \\
    & [$\unit[]{^{\circ}C}$] & [$\unit[]{^{\circ}C}$] & [$\unit[]{s}$] & [$\unit[]{s}$] \\
\hline \hline
1 & 10  & 35 &  60 & 1200 \\
2 & 15  & 30 &  60 & 1200 \\
3 &  5  & 40 &  60 & 1200 \\
4 & 17  & 33 & 120 & 1200 \\
5 & 10  & 35 &  90 &  900 \\
\hline \hline
\end{tabular}
\end{center}
\end{table}

\begin{table*}
  \caption{Synchronization quality in a multi-hop scenario for \textit{S}$_1$, \textit{NN}$_{link}$ (i.e, \textit{NN} trained for every hop), and \textit{NN}$_{gen_A}$ and \textit{NN}$_{gen_B}$ (i.e., the same \textit{NN} for every hop).}
  \label{tab:multi-hop2}
  \small
  \begin{center}
    \begin{tabular}{l|c||c|ccc||c|ccc|ccc|ccc}
      & Hop & & \multicolumn{3}{c||}{\textit{S}$_1$} & & \multicolumn{3}{c}{\textit{NN}$_{link}$} & \multicolumn{3}{c}{\textit{NN}$_{gen_A}$} & \multicolumn{3}{c}{\textit{NN}$_{gen_B}$} \\
      &   & $K$  & $\sigma$ & $p_{99.9}$ & Max & $K$ & $\sigma$ & $p_{99.9}$ & Max & $\sigma$ & $p_{99.9}$ & Max & $\sigma$ & $p_{99.9}$ & Max \\

      \hline \hline
      \multirow{5}{*}{\begin{sideways}WiFi ($sw^{WiFi}$)\end{sideways}}
 &1 &  20 & 0.207 &  0.717 &  1.184 &  60 & 0.186 &  0.761 &  1.202  & 0.183 &  0.772 &  1.192 &  0.198 &  0.832 &  1.262\\
 &2 &  20 & 0.400 &  1.339 &  2.103 &  60 & 0.280 &  1.090 &  1.669  & 0.330 &  1.245 &  1.805 &  0.308 &  1.136 &  1.988\\
 &3 &  20 & 0.635 &  2.369 &  3.946 &  60 & 0.432 &  1.553 &  2.413  & 0.528 &  2.180 &  3.045 &  0.477 &  1.862 &  3.176\\
 &4 &  20 & 0.855 &  3.060 &  5.403 &  60 & 0.493 &  1.673 &  2.359  & 0.711 &  2.802 &  5.030 &  0.561 &  2.093 &  4.039\\
 &5 &  20 & 1.110 &  3.846 &  6.415 &  60 & 0.600 &  2.006 &  2.734  & 0.946 &  3.854 & 11.052 &  0.668 &  2.567 &  4.972\\
        \hline \hline

        \multirow{5}{*}{\begin{sideways}WSN ($sw^{WSN}$)\end{sideways}}
 &1 &  40 & 0.633 &  2.205 &  3.349 & 60 &  0.647 &  2.794 &  3.479 &    0.623 &  2.267 &  3.104 &  0.676 &  2.328 &  3.018 \\
 &2 &  40 & 1.200 &  4.001 &  5.658 & 60 &  1.112 &  4.270 &  5.754 &    1.080 &  3.926 &  5.254 &  1.162 &  4.049 &  5.110 \\
 &3 &  40 & 1.947 &  7.111 & 10.200 & 60 &  1.561 &  5.290 &  7.314 &    1.860 &  7.089 &  9.176 &  1.720 &  6.020 &  7.786 \\
 &4 &  40 & 2.630 &  9.347 & 13.158 & 60 &  1.967 &  6.478 &  8.460 &    2.390 &  9.009 & 11.016 &  2.105 &  7.358 &  9.190 \\
 &5 &  40 & 3.440 & 11.831 & 16.825 & 60 &  2.355 &  7.626 & 10.500 &    2.999 & 11.198 & 13.862 &  2.484 &  8.744 & 11.163 \\
        \hline \hline

    \end{tabular}
    \\ All values are expressed in $\mu s$.
  \end{center}
\end{table*}

Table \ref{tab:multi-hop2} reports the experimental results. In addition to the $p_{99.9}$, also the standard deviation ($\sigma$) and the maximum (Max) of the absolute value of the synchronization error have been provided.

The three methods based on \textit{NN}, namely \textit{NN}$_{link}$, \textit{NN}$_{gen_A}$ and \textit{NN}$_{gen_B}$, have been compared with \textit{S}$_1$.

The first column of results of Table \ref{tab:multi-hop2} regards \textit{S}$_1$. For this experiment, the values $K^*$ obtained for the first hop (i.e., $20$ for $sw^{WiFi}$ and $40$ for $sw^{WSN}$) have been fixed and used for the remaining four hops. The symbol $K^*$ has been substituted in the result table with the symbol $K$, because the optimality of its value only applies to the first hop.

All the statistical indexes worsen with the distance from the root. Specifically, in the case of $\sigma$, the increase in the hop number is almost constant. The values of these constants are $\unit[0.201]{\mu s}$ for $sw^{WiFi}$ and $\unit[0.702]{\mu s}$ for $sw^{WSN}$. Roughly speaking, the value of $\sigma$ doubles every hop. A similar behavior affects $p_{99.9}$ and Max. Their trends with the distance are less deterministic because they are more affected by rare events.

In all the \textit{NN} methods presented, $K$ must be chosen large enough in order to obtain timestamps jitter compensation; it should be greater than the values $K^*$ obtained for \textit{S}$_1$ and for all the possible combinations of timestamps jitters and temperature variations. Fixing $K=60$ meets the requirement (values of $K$ greater than $60$ lead to similar results because the \textit{NN} is not too sensitive to changes of $K$).

The \textit{NN}$_{link}$ method is the best from the point of view of performance because the resulting NNs are specialized to cope with the specific experimental conditions of each couples of nodes, the network channel and the environment. The results reported in the column \textit{NN}$_{link}$ of Table \ref{tab:multi-hop2} confirm the ability of the \textit{NN} to outperform \textit{S}$_1$, and the benefit of using \textit{NN}$_{link}$ increases with the number of hops. Actually, the results regarding the synchronization between $N_1$ and $N_0$ in the first hop reflect exactly those reported for the same condition in Fig.~\ref{fig:exp_comparison}.

The training database of \textit{NN}$_{gen_A}$ has been derived by merging the training databases of five separate experiments in which the nodes $N_1$, $N_2$, $N_3$, $N_4$ and $N_5$ of Fig. \ref{fig:multihop} are directly connected with the reference clock $N_0$. In this case, the training patterns contain only the information regarding temperature variations and the in-node delays, but not the hop error. The resulting database contains $1500000$ patterns. As expected, all the performance indicators have worse performance than in the  \textit{NN}$_{link}$ case, but, excluding the maximum value of the $sw^{WiFi}$ condition for the hop number $5$, the synchronization quality of \textit{NN}$_{gen_A}$ is similar or outperforms the one of \textit{S}$_1$.

Under \textit{NN}$_{gen_B}$, the temperature conditions $5$, $4$, $3$, $2$, $1$ of Table \ref{tab:multi-hop1} have been used, in the reported inverted order from $5$ to $1$, to train the $5$ hops of the network presented in Fig.~\ref{fig:multihop}, respectively. The resulting database contains $1500000$ patterns.

The results reported in the last column of Table \ref{tab:multi-hop2} show that \textit{NN}$_{gen_B}$ performance is very closed to that of \textit{NN}$_{link}$. This tells us two things: that the use of a generalized \textit{NN} is possible because results are close enough to the optimum represented by \textit{NN}$_{link}$, and that the knowledge of the hop error is important because it drives to a better performance.

\subsection{Testbed}
Implementation and validation of synchronization protocols in real devices deserve specific attention. Results reproducibility is the first issue to be considered. Some experimental conditions may be difficult to be reproduced (i.e., temperature variations, in-node latencies or packet losses) as well as experiments duration that may be limited. The interested reader is referred to \cite{Cen15b} for details on reproducibility issues. How to measure the synchronization quality is another concern. In \cite{Cen15a}, the generation of two synchronous signals is used, together with an external device to check their time differences. Unfortunately, real devices add variable jitters when they trigger an actuation. In order to minimize this jitter, a hard real-time implementation is required. Moreover, a temperature-controlled environment may be also used to analyze temperature variations (see, e.g., \cite{Xu16}).

Some preliminary results of the \textit{NN} on the real implementation of \cite{Cen15a} are summarized here. Besides the fact that temperature variations are lower than the ones considered in the simulations (due to the heating system of the room considered), we obtained interesting results that confirm the applicability of the approach. The RBIS protocol \cite{TII-sync-industrial} is used to acquire two databases with respect to working days \cite{Cen15b} and weekend, respectively. The performance of the \textit{NN} is qualitatively comparable with the one obtained in simulations here when compared to the splines. An interesting result is that the \textit{NN} outperforms \textit{S}$_1$ for small value of $K$. This is useful in a WSN device because it reduces the size of the information vector and the inherent computation. Another promising issue relies on the fact that the \textit{NN} shows to be robust to loss of packets. This may open the door to outperforming other estimation approaches, specifically designed for intermittent observations, still being based on the Gaussian hypothesis. On the other hand, the \textit{NN} fails when trained on the week days and tested over the weekend. In order to prevent such a performance degradation, one may anticipate the working conditions of the \textit{NN}, in particular with respect to the temperature ranges to be addressed. This was validated by results not reported here for the sake of conciseness.

\section{Conclusions and Future Work} \label{sec:conclusions}
We have examined and discussed a neural estimation technique for the popular two-way timing message exchange synchronization protocol and for nodes affected by temperature variations and delay asymmetry.
The impacts of the delay knowledge and the presence of several hops in the network have been accurately analyzed. Numerical analysis reveals significant performance improvements over existing techniques (splines and \cite{Chaudhari}) under variable temperatures and different delay distributions. One of the most important outcomes is the robustness to increasing synchronization steps (high accuracy, independently to the number of timestamps used).

Future work includes different topics. The skew estimation and the robustness to loss of timestamps are currently under investigation. Other intriguing issues are: the impact of the knowledge of the temperature through observations, runtime retraining of the neural estimator to unexpected conditions, as well as the management of the multi-hop database under ``big data" paradigms. Preliminary results on a real implementation \cite{Cen15a} confirm the effectiveness of the proposed technique. A more in-depth validation on a number of real installations and environmental conditions is argument of future research as well.

\section*{Acknowledgment}
Computational resources were provided by HPC@POLITO, a project of Academic Computing within the Department of Control and Computer Engineering at the Politecnico di Torino (http://www.hpc.polito.it).

\appendix

\section{} \label{sec:AppendixA}

\begin{figure}[t]
\hskip-0.5in
\centering
\includegraphics[width=7.5cm]{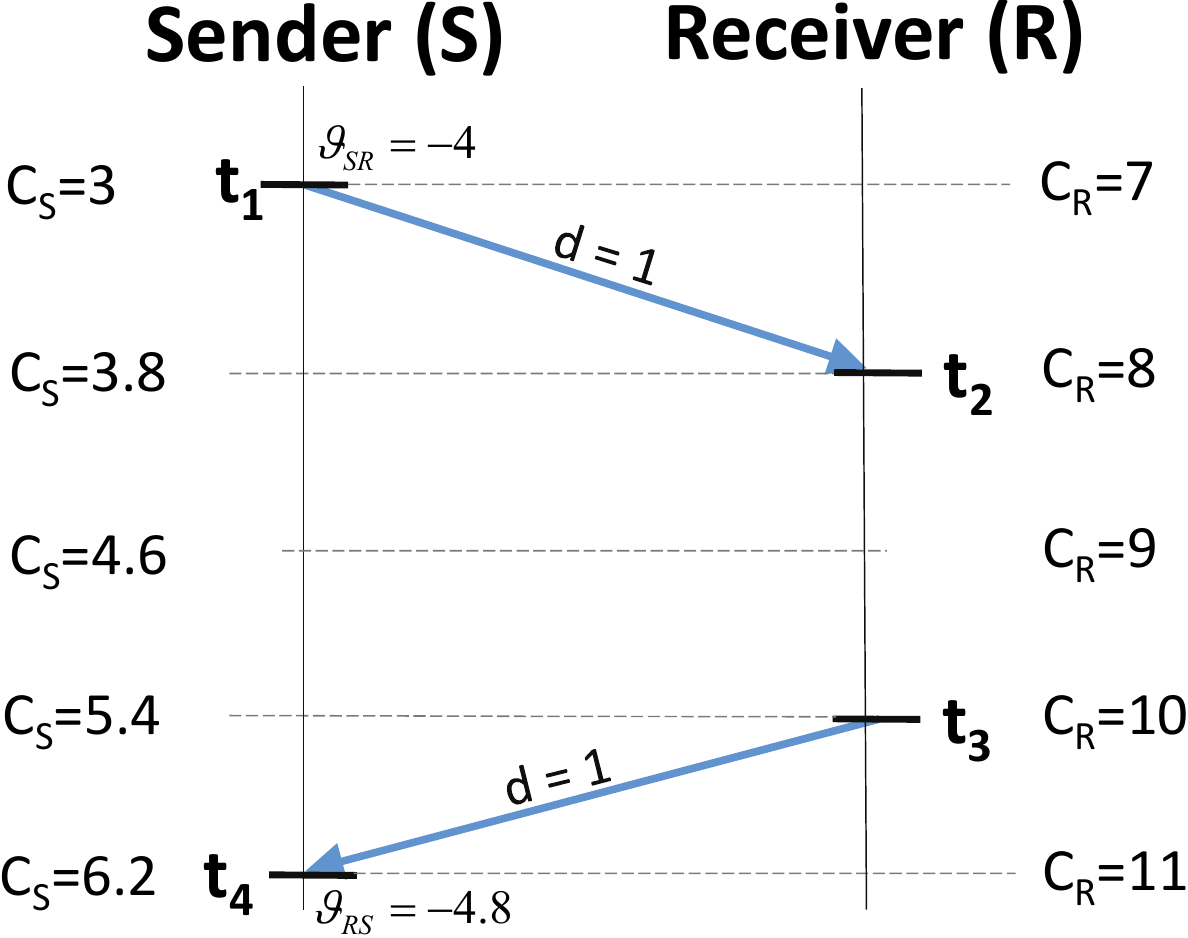}
\caption{Example of application of the measurement equations.}
\label{fig:delays_example}
\end{figure}

An example for explaining the measurement equations (\ref{eq:measure1})-(\ref{eq:measure3}) of subsection \ref{sec:measurementequations} has been reported in Fig.~\ref{fig:delays_example}. The index $k$ has been removed for the sake of simplicity and because only one message exchange is analyzed. The quantity $C_R$ is the free-running clock register of the \emph{receiver} node, which holds the reference time. Instead, $C_S$ (i.e., the free-running clock register of the \emph{sender} node) is updated with a lower frequency than $C_R$. The timestamps exploited for synchronization are: $t_1=3,\ t_2=8,\ t_3=10,\ t_4=6.2$.
The value $\gamma$ has been initialized with the estimation of the skew performed in the previous $k-1$ synchronization step (i.e., $\hat \gamma=-0.25$).
The new estimation of the delay from (\ref{eq:measure1}) is:
\begin{eqnarray}
\hat d & = & \frac{(1-\hat \gamma) \cdot (t_4-t_1)-(t_3-t_2)}{2} = \label{eq:d_example} \\
       & = & \frac{1.25 \cdot (6.2-3)-(10-8)}{2} = \frac{4-2}{2} = 1 \nonumber
\end{eqnarray}

By using the value $\hat d$ computed in (\ref{eq:d_example}), the new estimations of the offsets and of the skew can be obtained from (\ref{eq:measure2}) and (\ref{eq:measure3}):
\begin{eqnarray}
\hat \theta_{SR} & = & t_1 - (t_2-d) = 3-8+1 = -4  \\
\hat \theta_{RS} & = & t_4 - (t_3+d) = 6.2-10-1 = -4.8 \nonumber \\
\hat \gamma & = & 1 - \left( \frac{t_3+d-(t_2-d)}{t_4-t_1} \right) = 1 - \left( \frac{11-7}{6.2-3} \right) = \\
            & = & 1 - \frac{4}{3.2} = 1 - 1.25 = -0.25 \nonumber
\end{eqnarray}

\section{} \label{sec:AppendixB}
We demonstrate here that $1$-st order regression applied to timestamps is independent to the delay correction outlined in subsection \ref{sec:delaycorrection}.

The parameters of the $1$-st order regression (\textit{S}$_1$) interpolating a set of two-dimensional points $(X_k, Y_k), k=1,...,K$, i.e., slope $m$ and y-intercept $q$, are:
\begin{equation} \label{eq:retta}
m=\frac{\sum_{k=1}^K X_k Y_k-\frac{1}{K}\sum_{k=1}^K X_k \sum_{k=1}^K Y_k}{\sum_{k=1}^K X_k^2-\frac{1}{K}(\sum_{k=1}^K X_k)^2}; q=\bar{Y}-m\bar{X}
\end{equation}
Two straight lines are compared, with and without delay correction ($d$). The timestamps without correction consist of the sequence $\{t^k_1, t^k_2, t^k_3, t^k_4\}, k=1,...,K$; let $m$ and $q$ be the resulting \textit{S}$_1$ parameters. The timestamps with correction are $\{t^{k}_1, t_2^{k^\prime}, t_3^{k^\prime}, t^{k}_4\}$, $k=1,...,K$, $t_2^{k^\prime}=t^k_2-d$, $t_3^{k^\prime}=t^k_3+d$ with $m^\prime$ and $q^\prime$ the corresponding \textit{S}$_1^\prime$ parameters. The two sets of parameters asymptotically converge (in the number of timestamps) to identical values. If $m \sim m^\prime$ (`$\sim$' means asymptotical equivalence), $q \sim q^\prime$ because the averages $\bar{Y}$ and $\bar{X}$ (in \ref{eq:retta}) are identical in the two cases. In order to show $m\sim m^\prime$, we substitute in (\ref{eq:retta}) the timestamps and consider the difference $\Delta_m=m^\prime-m$ as $K\rightarrow+\infty$. After some algebraic manipulations, we obtain:

\begin{equation} \label{eq:rettasuitimestamps}
\Delta_m=\frac{d \sum_{k=1}^K (t^k_4 - t^k_1)}{\sum_{k=1}^K ((t^k_1)^2 + (t^k_4)^2)-\frac{1}{K}(\sum_{k=1}^K (t^k_1 + t^k_4))^2}
\end{equation}
The $\Delta_m$ quantity tends to $0$ for the following reasons. The numerator is positive ($t^k_4>t^k_1, \forall k$) and grows linearly. The denominator defines a definite positive quadratic form of the timestamps in $\Re^{+K}$; the determinant of the corresponding Hessian matrix is $(2^k \cdot \frac{(k-1)^k-1}{k^k})>0, \forall k>2$.

This asymptotic behavior is measured in practice with a small number of timestamps. For example, in the $sw^{WiFi}$ condition with $\tau=\unit[1]{s}$ and $K=60$, $\Delta_m=1.8 \cdot 10^{-9}$ and $\Delta_q=q^\prime-q = -\unit[54]{ns}$\footnote{The value $d=d^{prop}+d^{rec}-d^{send}=\unit[2.16]{\mu s}$ has been obtained with $d^{rec}=\mu^{rec}=\unit[7.23]{\mu s}$, $d^{send}=\mu^{send}=\unit[5.4]{\mu s}$ and $d^{prop}=\unit[334]{ns}$. $d^{prop}$ is the time the light takes to cover a distance of $\unit[100]{m}$. The quantity $d_R^{node}$ has been set to $\unit[0.5]{s}$. Values larger than $\unit[0.5]{s}$ lead to higher differences ($\Delta_m, \Delta_q$) and worse synchronization accuracy, the opposite holds for values smaller than $\unit[0.5]{s}$.}. The difference of the synchronization error between \textit{S}$_1^\prime$ and \textit{S}$_1$ is $\unit[54]{ns}$. By doubling the number of timestamps (i.e., $K=120$) the difference is halved, i.e., it amounts to $\unit[27]{ns}$, with $\Delta_m=4.5 \cdot 10^{-10}$ (an order of magnitude lower than with $K=60$) and $\Delta_q= -\unit[27]{ns}$.

\section{} \label{sec:AppendixC}
The asymmetry error outlined in subsection \ref{sub:exp2} can be a-priori computed by considering the propagation delays between the sender and the receiver nodes ($\overline d_{SR}^{path}$), and the one in the opposite direction ($\overline d_{RS}^{path}$):

\begin{eqnarray}
\overline d_{SR}^{path} & = & \overline d_{SR}^{prop}+\overline d_R^{rec} \cdot n_{\mu}-\overline d_S^{send} = \nonumber \\
                      & = & 0+\unit[7.23]{\mu s} \cdot 1.5-\unit[5.4]{\mu s}=\unit[5.45]{\mu s} \nonumber \\
\overline d_{RS}^{path} & = & \overline d_{RS}^{prop}+\overline d_S^{rec}-\overline d_R^{send}\cdot n_{\mu} = \nonumber \\
                      & = & 0+\unit[7.23]{\mu s}-\unit[5.4]{\mu s} = \unit[-0.87]{\mu s} \nonumber
\end{eqnarray}

As a consequence, the systematic error on the estimation of the propagation delay is:

\begin{eqnarray} \label{eq:calibration}
\overline \epsilon^{\check d(k)} = \frac{\overline d_{SR}^{path}-\overline d_{RS}^{path}}{2} = \frac{\unit[5.45]{\mu s}-\unit[0.87]{\mu s}}{2} = \unit[3.16]{\mu s} \label{eq:asym}
\end{eqnarray}

\bibliographystyle{elsarticle-num}
\bibliography{ADHOC-D-16-174}

\vspace{4cm}
\parpic{\includegraphics[width=1in,clip,keepaspectratio]{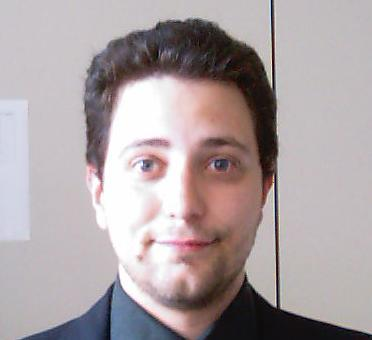}}
\noindent {\bf Maurizio Mongelli} got his Ph.D. at the University of Genoa in 2004. During the PhD and in the subsequent years, he worked on Quality of Service for military networks and Ethernet resilience for Selex Communications. He spent three months working on a project on satellite emulator systems at the German Aerospace Centre in Munich, Germany. He is co-author of over 70 scientific works, including international journals, conferences and patents. His main research activity concerns control of networks, machine learning and cybersecurity.

\parpic{\includegraphics[width=1in,clip,keepaspectratio]{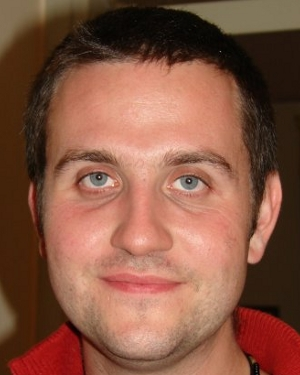}}
\noindent {\bf Stefano Scanzio} received the Laurea and Ph.D. degrees in computer science from Politecnico di Torino, Italy, in 2004 and 2008, respectively. From 2004 to 2009, he was involved with the Politecnico di Torino in research on speech recognition and classification methods and algorithms.

Since 2009, he was with the National Research Council of Italy (CNR), where he is a Tenured Technical Researcher with the IEIIT institute. He teaches several courses on computer science. He has authored/co-authored several papers in international journals and conferences in the area of synchronization protocols, industrial communication systems and real-time networks.

\end{document}